\documentclass[12pt,reqno]{article}
\usepackage[utf8]{inputenc}
\pdfoutput=1
\usepackage{setspace}
\usepackage{slashed}
\usepackage{graphicx,url}
\usepackage{chet}
\usepackage{booktabs}
\usepackage{amsmath,hyperref,amssymb}
\usepackage{color}
 
\usepackage{epsfig} 
\def\half{{1 \over 2}}

\def\ep{\epsilon}

\def\Or[#1]{{\text{O}}\left({#1}\right)}
\def\dotl[#1,#2]{\left\langle #1, #2 \right\rangle}
\def\dotlb[#1,#2]{[ #1, #2 ]}
\def\dotp[#1,#2]{(#1) \cdot (#2)}
\def\aff[#1,#2]{\hat{#1}(#2)}
\def\n4sym{{\cal N}=4 SYM}
\def\>{\rangle}
\def\<{\langle}
\def\weight[#1,#2,#3]{\{(#1),#2,#3\}}
\def\ads[#1]{$\text{AdS}_{#1}$}

\newcommand{\ba}{\begin{eqnarray}}
\newcommand{\ea}{\end{eqnarray}}
\newcommand{\be}{\begin{eqnarray}}
\newcommand{\ee}{\end{eqnarray}}
\newcommand{\bq}{\begin{equation}}
\newcommand{\eq}{\end{equation}}
\newcommand{\benn}{\begin{equation*}}
\newcommand{\eenn}{\end{equation*}}
\newcommand{\bi}{\begin{itemize}}  
\newcommand{\ei}{\end{itemize}}

\newcommand{\Acal}{{\mathcal A}} 

\newcommand{\Ccal}{{\mathcal C}}

\newcommand{\Ical}{{\mathcal I}}

\newcommand{\Lcal}{{\mathcal L}}
\newcommand{\Mcal}{{\mathcal M}}
\newcommand{\Ncal}{{\mathcal N}}
\newcommand{\Ocal}{{\mathcal O}}

\newcommand{\Pcal}{{\mathcal P}}

\newcommand{\Tcal}{{\mathcal T}}

\newcommand{\CI}{{\cal I}}
\newcommand{\CL}{{\cal L}}

\newcommand{\CO}{{\cal O}}

\newcommand{\nn}{\nonumber}

\newcommand\oo\infty
\newcommand\s\sigma
\newcommand\de\delta
\newcommand\De\Delta
\newcommand{\p}{\partial}
\newcommand\f\phi
\newcommand\g\gamma
\newcommand\x\times

\newcommand{\ra}{\rightarrow}

\newcommand{\fr}{\frac}
\newcommand{\comm}[2]{[#1,#2]}
\newcommand{\acomm}[2]{\{#1,#2\}}
\newcommand{\AdS}{\textrm{AdS}}
\newcommand{\CFT}{\textrm{CFT}}
\newcommand{\tfr}{\tfrac}

\newcommand{\sgn}{{\rm sgn}}
\newcommand{\eff}{\textrm{eff}}
\newcommand{\ET}{\textrm{ET}}
\newcommand{\LC}{\textrm{LC}}
\newcommand{\eps}{\varepsilon}
\newcommand{\Cmax}{\Ccal_{\max}}
\newcommand\lrpar{\raise .8ex\hbox{$^\leftrightarrow$} \hspace{-9pt}
\partial}

\newcommand\G{\Gamma}

\newcommand{\braket}[3]{\langle #1|#2|#3 \rangle}

\begin{document}

\preprint{CERN-TH-2018-030}
\title{Lightcone Effective Hamiltonians \\ and RG Flows}
\author{A. Liam Fitzpatrick$^1$, Jared Kaplan$^{2}$, Emanuel Katz$^1$, \\ Lorenzo G. Vitale$^1$, Matthew T.\ Walters$^{1,3}$}
\affiliation{
{\it $^1$ Department of Physics, Boston University, Boston, MA 02215} \\ 
{\it $^2$ Department of Physics and Astronomy, Johns Hopkins University, Baltimore, MD 21218 \\ 
Center for Quantum Mathematics and Physics (QMAP), 
UC Davis, CA 95616 \\
Stanford Institute for Theoretical Physics, Stanford University, Palo Alto, CA 94305} \\
{\it $^3$ CERN, Theoretical Physics Department, 1211 Geneva 23, Switzerland} \\
}

\abstract{
We present a prescription for an effective lightcone (LC) Hamiltonian that includes the effects of zero modes, focusing on the case of Conformal Field Theories (CFTs) deformed by relevant operators.    
We show how the prescription resolves a number of issues with LC quantization, including i) the apparent non-renormalization of the vacuum, ii) discrepancies in critical values of bare parameters in equal-time vs LC quantization, and iii) an inconsistency at large $N$ in CFTs with simple AdS duals.  We describe how LC quantization can drastically simplify Hamiltonian truncation methods applied to some large $N$ CFTs, and discuss how the prescription identifies theories where these simplifications occur. We demonstrate and check our prescription in a number of examples. 
}

\maketitle

\setcounter{tocdepth}{2}
\tableofcontents

\flushbottom


\section{Introduction and Summary}
\label{sec:intro} 

Quantum field theories (QFTs) can be defined as points along a renormalization group (RG)  flow between scale-invariant fixed points. One advantage of this formulation is that it does not make reference to a Lagrangian or a weak coupling expansion, but instead puts conformal field theory (CFT) fixed points front and center.  As methods for describing CFTs become more sophisticated, this formulation becomes increasingly useful as a practical computational tool.   

Of course the CFT endpoints are only half of the story, the other half being the dynamics of the RG flow. In many cases of interest, the flow is triggered by deforming the Hamiltonian by one of the relevant operators of the CFT,
\be
H = H_\CFT + V, \qquad V \equiv \lambda \int d^{d-1}x \, \Ocal_R(x).
\ee
The resulting theory can then be studied through non-perturbative Hamiltonian truncation techniques, which involve restricting the Hilbert space to a finite-dimensional subspace and numerically diagonalizing the truncated Hamiltonian exactly.
Yurov and Zamolodchikov were the first to derive the low-lying spectrum of QFT using the truncated spectrum approach \cite{Yurov:1991my,Yurov:1989yu}.
Recently, Hamiltonian truncation has been revived, in part thanks to several technical advancements that have improved 
the numerical predictivity of the method \cite{Hogervorst:2014rta,Rychkov:2014eea,Elias-Miro:2015bqk,Elias-Miro:2017xxf,Elias-Miro:2017tup}.
In the past few years Hamiltonian truncation has been applied with success to a variety of models, and to study several 
aspects of QFT, such as spontaneous symmetry breaking \cite{Coser:2014lla,Rychkov:2014eea,Bajnok:2015bgw}, 
scattering matrices \cite{Bajnok:2015bgw}, and quench dynamics~\cite{Rakovszky:2016ugs,Hodsagi:2018sul}.\footnote{A more comprehensive list of references can be found in \cite{James:2017cpc}.}

While many of the results and considerations in this paper should be generalizable to different UV bases, in this work we will 
focus on the particular implementation of \emph{conformal truncation} \cite{Katz:2013qua,Katz:2014uoa,Anand:2017yij,Katz:2016hxp}, which uses the eigenstates of the UV CFT Hamiltonian.  These states can be organized into representations of the conformal group, each of which is associated with a primary operator $\Ocal(x)$. Working in momentum space, we can write the states in the general form
\be
|\Ocal,\vec{P},\mu\> \equiv \int d^dx \, e^{-iP\cdot x} \Ocal(x)|0\> \qquad (\mu^2 \equiv P^2).
\ee
These states are characterized by an eigenvalue $\Ccal$ under the quadratic Casimir\footnote{This takes the familiar form $\Ccal = \Delta(\Delta-d) + \ell(\ell+d-2)$ in terms of operator dimensions and spins.} of the conformal group, spatial momentum $\vec{P}$, and invariant mass $\mu$. We can then truncate this basis by keeping  only those states with Casimir eigenvalue below a particular threshold, $\Ccal \leq \Cmax$. This approach allows us to study the resulting RG flow using only data from the original CFT (see~\cite{Katz:2016hxp} for a more detailed discussion). Specifically, the Hamiltonian matrix elements associated with the relevant deformation are constructed purely from CFT three-point functions.

However, in using this method (or any other Hamiltonian truncation approach), we must choose a quantization scheme in order to define both the basis states and the resulting Hamiltonian matrix elements.  The focus in this paper will be on lightcone (LC) quantization.  This quantization scheme involves using one spatial direction $x$ to define new LC coordinates $x^\pm \equiv \fr{1}{\sqrt{2}}(t \pm x)$. The Hilbert space is then defined on spacetime slices of fixed LC `time' $x^+$, with $x^-$ viewed as a spatial coordinate, along with the remaining transverse components $\vec{x}^\perp$.  The resulting Hamiltonian corresponds to the generator of LC time translations 
\be
H_\LC \equiv P_+.
\ee
Because equal-time (ET) quantization is better understood conceptually than LC quantization, it will be useful to know how to directly compare matrix elements in the two quantization schemes.  The comparison is given by the fact that LC quantization can be understood (and in fact was originally derived~\cite{Dirac:1949cp,Weinberg:1966jm,Bardakci:1969dv,Kogut:1969xa,Chang:1972xt}) as the infinite momentum limit of equal-time quantization.
 The individual matrix elements of the full theory (CFT + deformation) generically depend on the CFT invariant masses $\mu$, $\mu'$ of the two external states, as well as the overall spatial momentum $\vec{P}$. The corresponding LC quantization matrix elements can be obtained by taking the limit $|\vec{P}|\ra\infty$,\footnote{Part of the non-trivial content of this relation is that the matrix elements of $M^2$ in LC quantization are \emph{independent of the choice of reference frame} $P_-$. In other words, the matrix elements only depend on Lorentz invariant parameters, namely the original masses of the external states.  This is due to the fact that boosts simply rescale the LC Hamiltonian $P_+$ and can be used to completely factor out its $P_-$ dependence.  We show this explicitly in appendix \ref{app:LCvsET}. 
}
\be
\lim_{|\vec{P}|\ra\infty} M^2_\ET(\mu,\mu',\vec{P}) = M^2_\LC(\mu,\mu').
\label{eq:MsqETvsMsqLC}
\ee
where $M^2 = E^2 - \vec{P}^2 = 2P_+ P_- - \vec{P}_\perp^2$ is the mass-squared operator.  The main subtlety is that, due to truncation, the infinite momentum limit of the eigenvalues of $M_\ET^2$ are sometimes {\it not} the eigenvalues of $M^2_\LC$, and one may need to add new terms to the LC Hamiltonian to compensate for this effect.

In this paper we will develop a prescription that matches LC and ET Hamiltonians to all orders in the relevant deformation parameter.  This matching is non-trivial because LC quantization discards physical ``LC zero modes'' that are present in ET quantization.  This fact is responsible for many of the advantages of LC quantization, but also for several potential problems.  Our prescription was partly motivated by a desire to better understand when these advantages reflect true simplifications from LC quantization, and when they indicate that the LC treatment is missing a crucial aspect of the physics. 
Before we explain the prescription, we will review some of these advantages and disadvantages, and the precise relation between LC and ET quantization.
 Much of our discussion reviews well-known results~\cite{Klauder:1969zz,Wilson:1994fk,Brodsky:1997de}, but we will also emphasize a major advantage associated with large $N$ theories, which will be crucial to exploit in future applications.

\subsection{Advantages of Lightcone Quantization}
\label{sec:LCAdvantages}

\subsubsection*{Lack of Vacuum Renormalization}

Probably the most well-known simplification in LC quantization is the lack of vacuum renormalization. Physical states in LC quantization are required to have positive LC momentum,
\be
P_- \equiv \fr{1}{\sqrt{2}}(E - P_x) > 0,
\ee
leaving the vacuum as the \emph{unique} state with $P_- = 0$. So
conservation of LC momentum forbids any matrix elements which mix the vacuum with other states, 
and the interacting vacuum is naively the same as the Fock space vacuum,
\be
|\Omega\>_\LC = |0\>_\LC.
\ee 
This is advantageous as it eliminates the dependence of the physical state energies on the vacuum energy.\footnote{A noted closely related advantage is the potential absence of the ``orthogonality'' catastrophe, 
consisting in the fact that non-perturbative states in finite volume have exponentially small overlaps with perturbative 
states (see \cite{Elias-Miro:2017tup}, appendix A.1).}

As we will discuss, the correct interpretation of the statement that the vacuum state is not renormalized is a bit subtle. 
In particular, the fact that physical states in LC have $P_- >0$ really is the statement that states with $P_-=0$, which are present in ET, have been discarded, and our main focus will be on how to correctly reintroduce their effects.

\subsubsection*{Additional Selection Rules}

Positivity of LC momenta actually forbids a large class of Hamiltonian matrix elements, not just those involving the vacuum. For example, in the case where both external states are created by scalar operators, the matrix elements associated with the relevant deformation vanish when the scaling dimensions are related by an even integer,
\be
\<\Ocal,\vec{P},\mu|V_\LC|\Ocal',\vec{P}',\mu'\> = 0 \qquad (\De' = \De + \De_R + 2n).
\label{eq:DimReln}
\ee
We can clearly see that matrix elements involving the vacuum are merely a special case of this more general class, with $\De = 0$ and $\De' = \De_R$.

When the original CFT is a free theory, these lightcone selection rules forbid any process involving the creation of particles from the vacuum. For example, if our relevant deformation is a mass term, in ET quantization this operator would mix the one-particle state with all states containing odd numbers of particles. However, in LC quantization all of these matrix elements are set to zero, such that there is \emph{no} mixing between states with different particle numbers.

\subsubsection*{Major Simplifications for Large $N$ Theories}

The selection rule (\ref{eq:DimReln}) also simplifies large $N$ CFTs which are deformed by a relevant single-trace operator. As we will now explain, in lightcone quantization large $N$ RG flows appear to be fully determined by the planar OPE coefficients of (only) single-trace operators; in contrast equal-time quantization requires all planar OPE coefficients, including those of all multi-trace operators.

To understand this simplification, we must first briefly review the behavior of three-point functions in the large $N$ limit. For single-trace operators $\Ocal_i$, all OPE coefficients are suppressed by at least one power of $N$,\footnote{More precisely, this is our operational definition of ``$N$''.  For some large $N$ CFTs, this ``$N$'' will be a power of the rank of the underlying symmetry group. Also, note that all operators are normalized such that the two-point functions are $O(1)$. }
\be
\< \Ocal_i \Ocal_R \Ocal_j\> \sim \fr{1}{N}. 
\ee 
Consequently, the relevant deformation must scale linearly with $N$ in order to ensure that the resulting Hamiltonian matrix elements will be $O(1)$,
\be
V = N  \lambda \int d^{d-1} x \, \Ocal_R(x),
\ee
where the coefficient $\lambda$ is held fixed as $N \ra \infty$.

Matrix elements which mix these single-trace operators with generic multi-trace states $[\Ocal_i \cdots \Ocal_j]$ are suppressed by higher powers of $N$. For example, the three-point function involving a double-trace operator behaves as
\be
\< \Ocal_i \Ocal_R [\Ocal_j\Ocal_k]\> \sim \fr{1}{N^2}.
\ee 
The associated matrix element therefore vanishes in the infinite $N$ limit. However, there is a crucial exception, which is multi-trace operators $[\Ocal_i \dots \Ocal_R]$ involving the relevant operator $\Ocal_R$ itself. For instance,
\be
\< \Ocal_i \Ocal_R [\Ocal_i\Ocal_R] \> \sim 1 + \fr{1}{N^2},
\ee
where the leading $O(1)$ term corresponds to the known OPE coefficients for a generalized free field (GFF) \cite{Unitarity}. In ET quantization, the matrix elements of $V$ between states created by the single-trace operator $\Ocal_i$ and the double-trace operator $[\Ocal_i \Ocal_R]$ will therefore be $O(N)$.  Such contributions complicate the large $N$ limit and in particular prevent one from simply discarding matrix elements that vanish at infinite $N$, since diagonalization of the Hamiltonian can effectively multiply the $N$-suppressed matrix elements by the $N$-enhanced ones. In order to apply conformal truncation in ET quantization, we therefore need \emph{the full set of planar limit OPE coefficients} for the large $N$ CFT.

However, the leading GFF contributions to the Hamiltonian \emph{vanish} in LC quantization, precisely because the scaling dimensions of single- and double-trace operators are related by an integer at infinite $N$. Because of this, there are no longer any $O(N)$ terms in the Hamiltonian, which suggests that we can safely ignore any matrix elements which go to zero as $N\ra\infty$. Amazingly, this eliminates all matrix elements that mix single-trace states with multi-trace ones, which naively  means \emph{we only need the planar limit OPE coefficients of single-trace operators} in LC quantization. This is a much smaller set of data than the planar limit OPE coefficients of all operators, so the elimination of GFF matrix elements naively represents a striking simplification of the initial data that is required for conformal truncation.


\subsection{The Problem of Zero Modes}

Many of the above virtues have a corresponding dark side, associated with
LC zero modes. In LC quantization, any degrees of freedom with LC momentum $p_- = 0$ are non-dynamical and can therefore be removed from the Hilbert space. This removal of zero modes automatically follows from the definition of LC quantization as the infinite momentum limit of ET quantization, as all Hamiltonian matrix elements involving zero modes vanish as $|\vec{P}| \ra \infty$. In fact, it is precisely these vanishing matrix elements that lead to many of the simplifications discussed above.

This naively suggests that we can simply ignore zero modes in LC quantization, especially if we focus on states with finite $P_-$, and there are multiple examples in the literature where doing so apparently yields valid results (see \cite{Hiller:2016itl} for a recent review). However, as is well-known, there are also many cases where discarding zero modes leads to conceptual confusions and explicit, physical mistakes.   Some of the most important of these problems are the following:

\subsubsection*{Apparently Trivial Vacuum}

This problem is the flip-side of the advantage that the LC vacuum is apparently not renormalized and the vacuum energy naively receives no corrections, which suggests that there is no cosmological constant problem in LC quantization \cite{Brodsky:2008xu,Brodsky:2009zd}. However, this claim is clearly in conflict with both ET quantization results and standard Feynman-diagram perturbation theory \cite{Chang:1968bh,Yan:1973qg,Heinzl:2003jy,Herrmann:2015dqa,Collins:2018aqt}, and furthermore
leads to conceptual difficulties in the case of spontaneous symmetry breaking (SSB) \cite{Maskawa:1975ky,Tsujimaru:1997jt,Yamawaki:1998cy, Beane:2013ksa}.

\subsubsection*{Insensitivity to Tadpoles}

The problems associated with zero modes are not just limited to the vacuum, however.   An especially simple example is that of a scalar field theory deformed by a tadpole,
\be
V(\phi) \ra V(\phi) + \lambda\phi.
\label{eq:tadpole}
\ee
For typical $V(\phi)$, the addition of the tadpole shifts the mass and couplings in the theory, with observable consequences for the resulting spectrum and RG flow. Yet, the contribution of the tadpole to the action is purely a zero mode of $\phi$, which means it will have no effect on the Hamiltonian matrix elements if zero modes are not included. 

\subsubsection*{Incorrect Predictions for Simple Holographic Models}

While the above tadpole example (\ref{eq:tadpole}) may seem a bit special, there is a very similar problem which arises in deformations of large $N$ theories by single-trace operators. A simple toy example is a large $N$ CFT dual to an effective field theory in anti-de Sitter (AdS) with the bulk Lagrangian
\be
\Lcal_{\textrm{bulk}} = \half \p^\mu\phi \p_\mu\phi - \half m^2 \phi^2 - \fr{1}{4} \fr{g_4}{N^2} \phi^4.
\ee
If we deform this theory by the single-trace operator $\Ocal$ dual to the scalar field $\phi$, the arguments of the previous section suggest that the resulting RG flow  depends at leading order only on the single-trace three-point function $\<\Ocal \Ocal \Ocal\>$, since any contributions to the LC Hamiltonian from multi-trace operators vanish in the limit $N \ra \infty$.  However, it is clear from solving the bulk equations of motion for $\phi$ that the dynamics are sensitive to $g_4$ even in the infinite $N$ limit.

\subsubsection*{Discrepancy Between Bare Parameters in ET and LC}

If the lightcone Hamiltonian $V_\LC$ is defined as the infinite momentum limit of the equal-time Hamiltonian $V_\ET$, then naively the bare parameters (which are Lorentz invariant) associated with the relevant deformations should be the same in both quantization schemes. However, there are cases where the two schemes obtain \emph{different} mass eigenvalues when using the \emph{same} bare parameters \cite{Burkardt:1992sz,Burkardt:2016ffk}.

\subsubsection*{Obstacles to Integrating Out Zero Modes}

All of the foregoing difficulties with zero modes reduce to the same core problem: LC quantization is missing contributions to the Hamiltonian which are present in ET quantization. In some cases these contributions can safely be ignored, while in others they can't, with no clear a priori diagnostic for determining when and no systematic method for reintroducing the necessary effects. 

A natural strategy for trying to deal with the problem of the missing zero modes is to try integrating them out, rather than simply discarding them.  However, there are signs that in Discrete Lightcone Quantization (DLCQ) integrating out zero modes can lead to new strongly coupled interactions between the remaining modes \cite{Hellerman:1997yu}.


\subsection{Our Prescription}

In this work, we will describe how to overcome most of the problems described above, by proposing a general prescription for absorbing the effects of zero modes into a new effective LC Hamiltonian to all orders in the relevant deformation parameter. This proposal is formulated directly in terms of correlators of the UV CFT, and is thus not restricted to theories with known Lagrangian descriptions. The prescription is essentially a matching procedure, where we construct an effective lightcone Hamiltonian $H_{\rm eff}$ for the theory without any zero modes that reproduces all correlation functions of the theory in the presence of zero modes.  To connect the Hamiltonian to correlators, we define it in terms of the LC unitary evolution operator $U$ as
\be
H_{\rm eff} \equiv \lim_{x^+ \rightarrow 0} i\partial_+ U(x^+),
\label{eq:Heff}
\ee
where the evolution operator is constructed from the naive lightcone Hamiltonian $V_\LC$ (i.e.~without including the effects of zero modes). Through the Dyson series for $U(x^+)$, the matrix elements for $H_\eff$ can be written in terms of correlators of the original CFT involving multiple insertions of $V_\LC$. We thus ``integrate out'' the zero modes by embedding their contributions to higher-point functions into $H_{\rm eff}$ via eq.~\eqref{eq:Heff}.

We can therefore understand the effects of zero modes  by looking at higher-point correlation functions of the general form
\be
\<\Ocal,\vec{P},\mu|\Tcal\{\Ocal_R(x_1) \cdots \Ocal_R(x_n)\}|\Ocal',\vec{P}',\mu'\>.
\ee
If these correlators are regular as $x_{ij}^+ \ra 0$, then all the higher-point contributions to~\eqref{eq:Heff} will vanish, reducing our prescription to the standard definition of Hamiltonian matrix elements in terms of three-point functions. However, as we will show, there are cases where these correlators include effects that do not have a spectral representation in LC quantization, which leads to factors of $\de(x_{ij}^+)$. This singular behavior is picked up by our prescription for $H_\eff$, resulting in corrections to the naive Hamiltonian.

We demonstrate that our conjectured prescription:
\bi
\item results in a non-zero contribution to the vacuum energy,
\item reproduces the shifts in masses and couplings due to tadpole deformations,
\item includes the effects of multi-trace operators on large $N$ RG flows,
\item explains the discrepancy between bare parameters in ET and LC  perturbatively, 
\item automatically integrates out non-dynamical fields. 
\ei 
In fact, in {\it most} CFTs, $H_{\rm eff}$ will not get any contributions from zero modes aside from the vacuum energy.  
In addition, we will provide evidence that in many theories where $H_{\rm eff}$ does get contributions from zero modes, those contributions simply shift the bare parameters in the original theory. 

Our prescription for matching the LC and ET Hamiltonians is perturbative in the relevant deformation parameters, and can fail non-perturbatively.  We will discuss an explicit example of this failure in section \ref{sec:phi4}.  One might nevertheless hope that knowing the perturbative matching can still be useful for understanding qualitative or even quantitative aspects of the non-perturbative matching.\footnote{Moreover, the prescription is still non-perturbative in terms of the parameters of the UV CFT, which can for instance include the gauge coupling if the UV CFT is a gauge theory.}  We leave a more detailed analysis of such non-perturbative matching effects for future work.

Another remaining important open question is how to get the vacuum structure correct in cases of spontaneous symmetry breaking (SSB).  It will be important to determine whether the LC methods we adopt in this approach are sufficient for correctly reproducing the broken phase of SSB, or whether they must be supplemented with additional inputs.  For instance, one concern is that there is no SSB in finite volume, since mixing between different vacua lead the true ground state to be a superposition of the infinite volume symmetry-breaking vacua.
Although formally we work in a framework where the volume is infinite, one may worry that the truncation itself  causes the system to behave more like finite volume for SSB effects.  At a more technical level, we will see that our prescription applied to the theory of a scalar field perturbed by a source term $\CL \supset J \phi$ simply generates the terms in the Hamiltonian produced by expanding around the new shifted vacuum. However, there are generally multiple local extrema of the potential, and it is not clear if the correct choice needs to be put in by hand in the LC treatment, or whether it can be selected dynamically.\footnote{For an illuminating physical picture of how this can occur, see \cite{Rozowsky:2000gy}.  Essentially the same issue of whether or not a fundamental field vev must be set by hand in a LC treatment arises in holographic models we consider, where a necessary input to the boundary value problem for a bulk field $\phi$ profile is the boundary vev $\< \CO\>$ of its boundary dual operator. } A useful concrete check would be to compute the spectrum of the theory in the broken phase, for instance in $\lambda \phi^4$ theory, and see if the result correctly reproduces the spectrum of fluctuations around one of the $\mathbb{Z}_2$-breaking vacua.  

This paper is organized as follows.  In section \ref{sec:Discontent}, we describe in more detail some of the problems with LC quantization mentioned above.  In section \ref{sec:Prescription}, we present our prescription for the effective LC Hamiltonian $H_{\rm eff}$, together with a quantitative diagnostic test in momentum space for whether or not the prescription generates new contributions to $H_{\rm eff}$ in a given theory. In section \ref{sec:Applications}, we demonstrate how the prescription works in a number of applications.  One of the main applications is to $\lambda \phi^4$ theory in 2d, where the prescription reduces to a previous prescription due to Burkardt \cite{Burkardt:1992sz}.  Our numeric results are consistent with the conjecture that the prescription works to all orders in perturbation theory, but indicate that it fails non-perturbatively.  In section \ref{sec:discussion}, we conclude with a discussion of future directions.


\section{Lightcone Quantization and its Discontents}
\label{sec:Discontent}

In this section, we provide a more detailed discussion of some of the problems with LC quantization listed in the introduction.  We begin with a definition of lightcone zero modes in section \ref{sec:DefinitionZeroModes} and a discussion of their role in CFTs.  Then we briefly review the discrepancy between ET and LC bare parameters in section \ref{sec:BareCoupling}.  Finally,  in section \ref{sec:Problem} we discuss a new problem that has great importance for the application of Hamiltonian truncation to large $N$ gauge theories.

\subsection{CFT Definition of Zero Modes}
\label{sec:DefinitionZeroModes}

Free and large $N$ CFTs have a Fock space description.  In these cases lightcone zero modes can be easily  identified as the states where one or more  Fock space modes have vanishing lightcone momentum $p_- = 0$. 

However, we would like to have a more general, non-perturbative definition of zero modes, which can be applied to any CFT. Here, we construct such a definition in terms of the associated Hamiltonian matrix elements, which  will allow us to easily demonstrate why these contributions naively vanish in LC quantization (or equivalently, in the infinite momentum limit of ET quantization). The derivation of this result will be somewhat schematic, with a more careful proof presented in appendix~\ref{app:MEfromC}.

In the standard formulation of conformal truncation, Hamiltonian matrix elements associated with a relevant deformation $\Ocal_R$ are defined as the Fourier transform of CFT three-point functions,
\be
\<\Ocal,\vec{P},\mu|V|\Ocal',\vec{P}',\mu'\> \equiv \lambda \int d^dx_1 \, d^{d-1}x_2 \, d^dx_3 \, e^{i(P\cdot x_1 - P'\cdot x_3)} \<\Ocal(x_1) \Ocal_R(x_2) \Ocal'(x_3)\>.
\ee
As is well-known, these three-point functions are completely fixed by conformal symmetry, up to overall constants corresponding to OPE coefficients. For the case where the two external operators are scalars, these correlators take the form
\be
\<\Ocal(x_1)\Ocal_R(x_2)\Ocal'(x_3)\> = \fr{C_{\Ocal\Ocal'\Ocal_R}}{x_{12}^{\De+\De_R-\De'} x_{23}^{\De'+\De_R-\De} x_{13}^{\De+\De'-\De_R}}.
\ee

A useful formal trick for studying the universal kinematic structure of this three-point function (and thus the resulting Hamiltonian matrix element) is to pretend that the three operators are composites constructed from building blocks $A$, $B$, and $C$,
\be
\Ocal_R \equiv AB, \quad \Ocal \equiv AC, \quad \Ocal' \equiv BC,
\ee
where the scaling dimensions for these new operators are
\be
\De_A = \half (\De + \De_R - \De'), \quad \De_B = \half (\De' + \De_R - \De), \quad \De_C = \half (\De + \De' - \De_R).
\ee
Note that we are \emph{not} assuming that this CFT is free or has a large $N$ expansion. These building blocks are merely a means of representing the kinematic structure of three-point functions. Because these operators are fictitious, their dimensions are not necessarily bounded from below due to unitarity and can therefore even have negative scaling dimension.

Using these building blocks, the kinematic structure of this three-point function simply becomes a product of two-point functions,
\be
\<\Ocal(x_1)\Ocal_R(x_2)\Ocal'(x_3)\> \propto \<A(x_1)A(x_2)\>\<B(x_2)B(x_3)\>\<C(x_1)C(x_3)\> = \fr{1}{x_{12}^{2\De_A} x_{23}^{2\De_B} x_{13}^{2\De_C}}.
\ee
Similarly, we can rewrite the Hamiltonian matrix element in terms of the spectral densities of these fictitious operators,
\be
\<\Ocal,\vec{P},\mu|V|\Ocal',\vec{P}',\mu'\> = \lambda \, C_{\Ocal\Ocal'\Ocal_R} \, \de^{d-1}(P - P') \int d^dp \, \rho_A(P-p) \rho_B(P'-p) \rho_C(p).
\label{eq:Triangle}
\ee

\begin{figure}[t!]
\centering
\includegraphics[width=.5\linewidth]{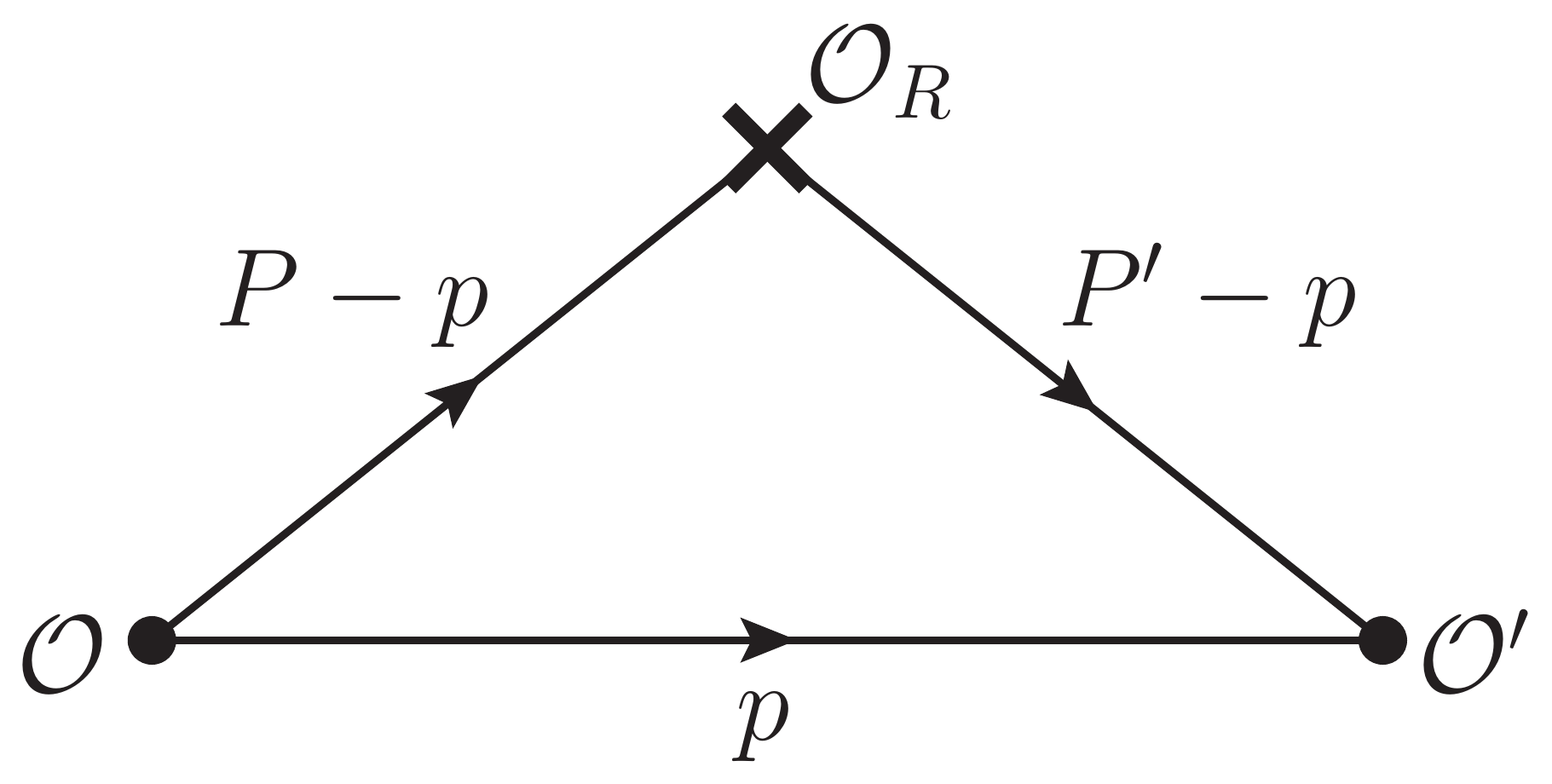}
\caption{Triangle diagram associated with the matrix element $\<\Ocal|V|\Ocal'\>$, demonstrating the flow of momentum. Each leg of the triangle can be thought of as the momentum space two-point function of a fictitious ``building block'' operator.  Lightcone zero modes are defined to be contributions where one of the legs has vanishing lightcone momentum. }
\label{fig:Triangle}
\end{figure}

We can therefore think of this matrix element as an integral over the momentum space two-point functions of the building blocks, as shown in the schematic ``triangle diagram'' in figure~\ref{fig:Triangle}. These diagrams are useful in picturing the flow of momentum in the associated matrix element. The vertices correspond to the insertions of the physical local operators, while the legs correspond to the internally propagating building blocks.

We can now use this representation of the matrix elements \emph{to define the contribution of zero modes in a general CFT as any diagram where one of the legs has zero lightcone momentum} (i.e.\ $p_- = 0$ or $p_- = P_-$). Note that this definition naturally encompasses the more familiar case of free field theory, where the legs of the triangle diagram simply correspond to one or more internally propagating Fock space modes.

Generically, we expect such contributions to be a measure-zero part of the full integral. However, in the special case where the dimensions of the three external operators are related by an non-negative even integer,
\be
\De' = \De + \De_R + 2n,
\label{eq:DimRelation}
\ee
one of the fictitious building blocks obtains a non-positive integer scaling dimension,
\be
\De_A = -n, \quad \De_B = \De_R+n, \quad \De_C = \De+n.
\ee
For this special case, the associated spectral density is given by a derivative of the Dirac delta function. For example, in $d=2$ we obtain
\be
\rho_A(P-p) = \de^{(n)}(P_+-p_+) \de^{(n)}(P_- - p_-) \qquad (\De_A = -n),
\ee
with similar expressions in higher dimensions. The spectral density for $A$ therefore fixes the internal momentum $P_- - p_- = 0$, such that \emph{only zero modes contribute}.

Focusing specifically on the lightcone momentum dependence of the Hamiltonian matrix element (and suppressing all other factors), we can then obtain\footnote{Here we've used the fact that CFT spectral densities scale as $\rho_\Ocal(p) \sim p^{2\De-d}$.} 
\be
\begin{split}
\<\Ocal,\vec{P},\mu|V|\Ocal',\vec{P}',\mu'\> &\propto \int dp_- \, \de^{(n)}(P_- - p_-) \, (P'_- - p_-)^{\De_R+n-\fr{d}{2}} p_-^{\De+n-\fr{d}{2}} \\
&\propto (P_- - P'_-)^{\De_R-\fr{d}{2}}.
\label{eq:SmallOVO}
\end{split}
\ee
These zero mode matrix elements are thus set by the difference in total lightcone momentum between the two external states. In LC quantization, conservation of momentum automatically sets this difference to zero, but we can also see that this difference vanishes in the infinite momentum limit of ET quantization,
\be
P_- - P'_- = \Big( \sqrt{\mu^2+P_x^2} - P_x \Big) - \Big( \sqrt{\mu'^2+P_x^2} - P_x \Big) \sim \fr{\mu^2 - \mu'^2}{2|P_x|} \ra 0 \quad (|P_x| \ra \infty).
\ee

For deformations with $\De_R > \fr{d}{2}$, we therefore find that \emph{all zero mode contributions vanish in lightcone quantization}. For deformations with $\De_R \leq \fr{d}{2}$, the story is somewhat more subtle. Rather than vanishing, the associated matrix elements are all IR divergent. We expect that one can regulate these divergences with some IR cutoff, and then take the limit $\Lambda_{\textrm{IR}} \ra 0$, so that all contributions from zero modes decouple from the resulting low-energy states and can be removed.\footnote{For more details in the particular case of a free theory, see~\cite{Katz:2016hxp}, where the decoupling of zero modes due to IR divergences naturally led to a rearrangement of the naive conformal basis into new ``Dirichlet'' states with no overlap with zero modes.}

Note that for free or large $N$ theories, these zero mode matrix elements precisely correspond to the case where one of the operators is a composite  built from the other two,
\be
\Ocal' = [\Ocal \Ocal_R]_n \equiv \Ocal \lrpar^{2n} \Ocal_R,
\ee
such that we can interpret these vanishing contributions as the creation of the $\Ocal_R$ degrees of freedom from the vacuum.

\subsection{Concrete Bare Parameter Discrepancies}
\label{sec:BareCoupling}

In some cases it had already been recognized in the literature that the ``naive'' light-cone Hamiltonian is missing contributions
that prevent a precise matching to equal time computations. In particular, in \cite{Burkardt:1992sz} (see also 
\cite{Burkardt:2016ffk}) it was pointed out that in a scalar theory, the effect of zero modes is to renormalize bare 
parameters on the lightcone.

For example, consider the deformation of free scalar field theory by both a mass term and a quartic interaction,
\be
V = \int d^{d-1}x \left( \half m^2 \phi^2 + \fr{1}{4!} \lambda \phi^4 \right).
\ee
As was demonstrated in~\cite{Burkardt:1992sz}, if we compute the one-particle mass perturbatively, we find that there are a class of Feynman diagrams which contribute in ET quantization but vanish in LC quantization, leading to a discrepancy in the resulting physical mass eigenvalues as a function of the bare parameters $m^2$ and $\lambda$. From the perspective of conformal truncation, these Feynman diagrams are constructed from intermediate Hamiltonian matrix elements which vanish on the lightcone. This discrepancy can in principle be fixed by shifting the LC bare mass relative to the ET value,
\be
m^2_\LC = m^2_\ET + \de m^2(\lambda),
\ee
where the coupling-dependent counterterm corresponds to resumming all diagrams which contribute in ET but not LC quantization. However, this fix may seem unsatisfying, as the counterterm must be introduced by hand, with no general prescription for determining when corrections are necessary.

In \cite{Burkardt:1992sz,Burkardt:1997bd}, it was argued by inspection of Feynman diagrams that the correct matching should be
\be
m^2_\LC = m^2_\ET + \frac{ \lambda}{2} \< \phi^2\>, 
\ee
where the vev of $\phi$ is evaluated in ET quantization. In section \ref{sec:phi4}, we will verify this formula explicitly to the first few orders in perturbation theory, and discuss its failure non-perturbatively.

\subsection{A Problem with Holographic Models}
\label{sec:Problem}

Simply discarding zero modes leads to incorrect predictions in a simple class of CFT models defined holographically using AdS Lagrangians.  The main point is quite simple:  AdS models can have contact interactions involving $n > 3$ bulk fields that have very important effects on RG flows, which are represented as non-trivial classical solutions in the bulk. But these interactions will be invisible if one only studies the OPE coefficients of single-trace operators (i.e.~3-pt interactions of bulk fields). At large $N$, this is in direct conflict  with the naive LC selection rule (\ref{eq:DimReln}), which implies that only single-trace data should affect RG flows.

The simplest explicit example includes a real scalar field in AdS with bulk\footnote{We will work with Poincar\'e patch coordinates
\be
ds^2 = \frac{-dz^2 + d x_d^2}{z^2}.
\ee
}
 Lagrangian
\be
\Lcal_{\rm AdS} = \frac{1}{2} (\partial \phi)^2  - \frac{1}{2} m^2 \phi^2 - \fr{1}{4} \frac{g_4}{N^2} \phi^4.
\ee
We are just going to work in the semi-classical limit at large $N$, so we will rescale the field $\phi \rightarrow N \phi$ to put the AdS Lagrangian in the form
\be
\CL_\AdS = N^2 \left( \frac{1}{2} (\partial \phi)^2 - \frac{1}{2} m^2 \phi^2  - \frac{1}{4 } g_4 \phi^4 \right).
\ee
We assume that $g_4>0$ for stability at $\phi \rightarrow \pm \infty$.  We will be interested in  deforming the boundary theory by the CFT operator $N \Ocal_R$ dual to $\phi$, and we take $m^2  = \ell_{\rm AdS}^{-2} \Delta_R (\Delta_R-d)$ negative, which corresponds to  $\Ocal_R$ being a relevant operator with $\Delta_R < d$.

The key point is that at infinite $N$, the quartic coupling $g_4$ does not affect any of the single-trace OPE coefficients. In fact, for this particular example, the $\mathbb{Z}_2$ symmetry of the AdS Lagrangian restricts all single-trace OPE coefficients to be exactly zero. Therefore, if the logic in the previous section is correct, $g_4$ cannot have any effect on the theory at infinite $N$, even after deforming by the relevant operator $ \Ocal_R$ in the boundary CFT Hamiltonian. In fact, the argument from the previous section would predict that there is \emph{no resulting RG flow}, since all Hamiltonian matrix elements for this deformation vanish at infinite $N$ in lightcone quantization. We will now demonstrate that there \emph{is} a resulting RG flow and $g_4$ {\it does} in fact affect the IR of the theory, so something in the previous section must have been too fast.   The  idea is very simple -- holographic RG flows involve solutions to the classical equations of motion for $\phi$, and $g_4$ will obviously affect these solutions.

Turning on the relevant deformation $ \Ocal_R$ in the boundary CFT corresponds in the bulk description to imposing a non-zero boundary condition for $\phi$:
\be
V = N \lambda \Ocal_R \quad \leftrightarrow \quad \phi(z) \stackrel{z\sim 0}{\sim}  \lambda z^{d-\Delta_R} + \alpha z^{\Delta_R} .
\ee
The second boundary value $\alpha$ can be determined dynamically once the bulk profile is known. To find the bulk profile, one imposes the bulk equations of motion for $\phi$. For $\phi$ constant in the boundary directions $x_d^\mu$, the bulk equation of motion is just
\be
\partial_z^2 \phi - \frac{d-1}{z} \partial_z \phi -\frac{m^2}{z^2} \phi = \frac{1}{z^2} g_4 \phi^3.
\ee
For any value of $\lambda$ and $\alpha$, there is a unique solution $\phi_{\rm cl}$ to this equation, and $\alpha$ is chosen to minimize the action $S[\phi_{\rm cl}]$ evaluated on this solution.  When $g_4=0$, the bulk theory is free and the equation of motion is easily solved by $\phi_{\rm cl} = \lambda z^{d-\Delta_R} + \alpha z^{\Delta_R}$ everywhere.  Substituting back into the action, 
\be
S_{g_4=0}[\phi_{\rm cl}] = \frac{N^2}{2} \left[ z^{d-2\Delta_R} \lambda^2 (d-\Delta_R) + z^{2\Delta_R-d} \alpha^2 \Delta_R \right]_{z_{\rm IR}}^{z_{\rm UV}}.
\ee
This action is minimized at $\alpha=0$.  

\begin{figure}[t!]
\begin{center}
\includegraphics[width=0.6\textwidth]{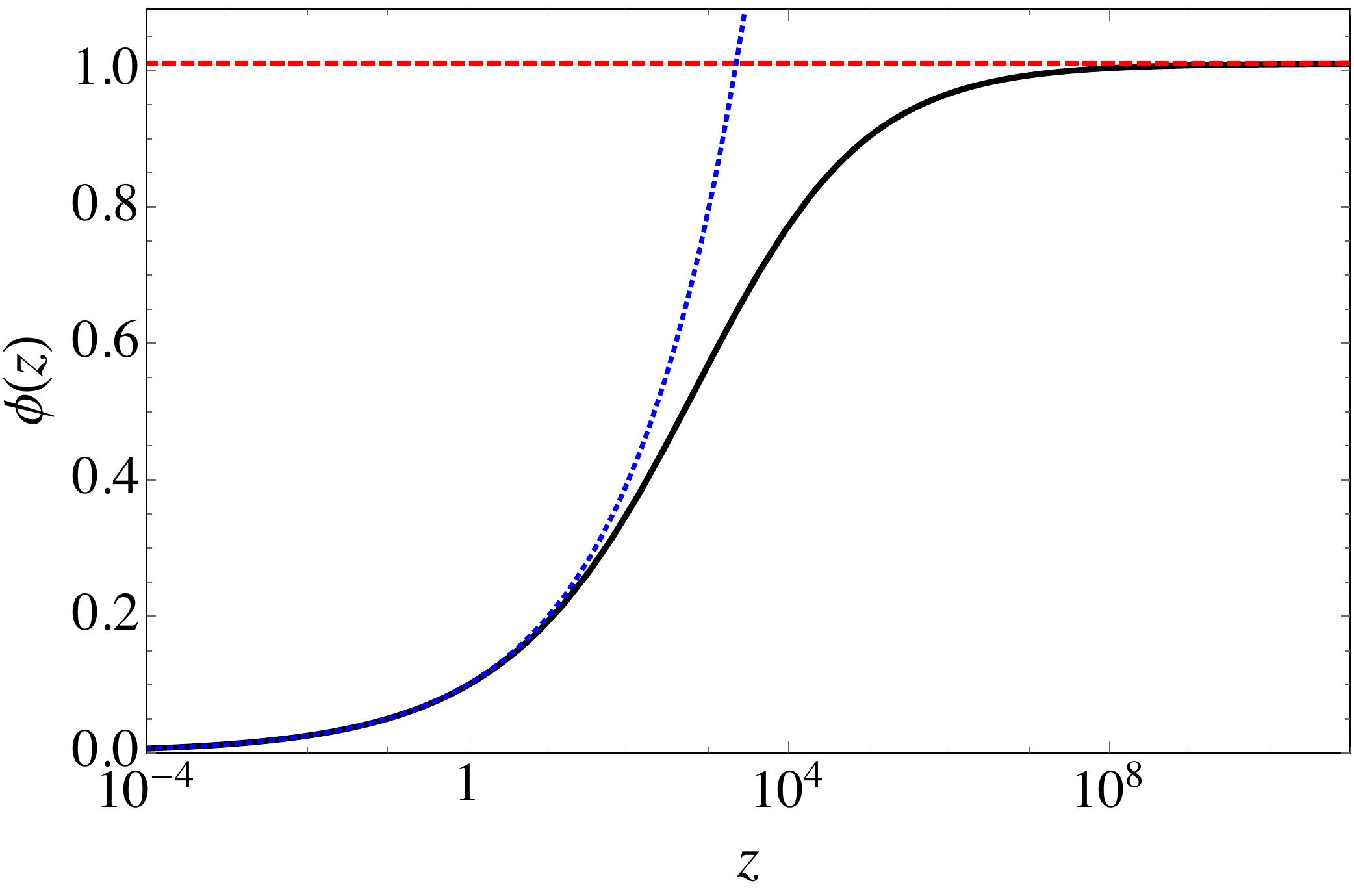}
\caption{Numeric solution of the bulk profile $\phi(z)$ in the toy model.  ({\it Black, solid}): exact numeric solution; ({\it red, dashed}): asymptotic value at $\phi = \sqrt{\frac{-m^2}{g_4}}$; ({\it blue, dotted}): free theory behavior $\phi = \lambda z^{d-\Delta_R}$.  Parameters are $\Delta_R = 1.7$, $\lambda =0.1$, $g_4 = 0.5$, $z_{\rm UV} = 10^{-4}$, $d=2$, all in units of $\ell_{\rm AdS}=1$. }
\label{fig:toybulksoln}
\end{center}
\end{figure}

However, the situation is qualitatively changed for any non-zero value of $g_4$. For $g_4>0$, the bulk equations of motion cannot be solved in closed form.  In appendix \ref{app:SUSY} we describe a supersymmetric version of this model where an analytic solution is possible, but the important qualitative points can be understood intuitively as follows.    Because $\Delta_R < d$, $\lambda z^{d-\Delta_R}$ grows as $z$ increases for any non-zero value of $\lambda$, and therefore the bulk term $g_4 \phi^4$ in the potential eventually becomes important for large enough $z$. The value of $\alpha$ that minimizes the action is the one that causes $\phi$ to asymptotically approach the minimum of its bulk potential $V_\AdS(\phi) = \frac{1}{2} m^2 \phi^2 + \frac{g_4}{4} \phi^4$ at large $z$.  In other words, since $m^2 < 0$, any non-zero $\lambda$ pushes $\phi$ away from the origin and, as it evolves into the bulk, it rolls down its potential to the true minimum at $\phi = \pm \sqrt{\frac{-m^2}{g_4}}$. A numeric solution exhibiting this behavior is shown in figure~\ref{fig:toybulksoln}.  

To determine the spectrum of the theory, one expands $\phi$ around the background solution $\phi_{\rm cl}$.   In the deep IR, the background $\phi_{\rm cl}$ is just a constant and so one can do this expansion analytically.  The fluctuations around $\phi_{\rm cl} = \pm \sqrt{\frac{-m^2}{g_4}}$ have a bulk mass of $V''_\AdS(\phi_{\rm cl}) = -2 m^2$. This mass corresponds to an IR dimension for $\Ocal_R$ of
\be
\Delta_{\rm IR} = 1 + \sqrt{1+ 2 \Delta_R(2-\Delta_R)} .
\ee
In the language of the CFT, turning on the relevant deformation triggers an RG flow from a UV CFT where $\Ocal_R$ has dimension $\Delta_R$ to an IR CFT where it has dimension $\Delta_{\rm IR}$.

To summarize the main point,  at large $N$ $g_4$ is completely invisible in the OPE coefficients of single-trace operators, yet from the bulk solution we see that the value of $g_4$ controls when the theory deformed by $\Ocal_R$ transitions from the UV behavior with dimension $\Delta_R$ to the IR behavior with dimension $\Delta_{\rm IR}$.   In fact, any term $\phi^n$  in the bulk potential with $n\ge 4$ is invisible to the single-trace OPE coefficient in the infinite $N$ limit, yet from the bulk perspective it is clear that they affect the IR of the theory.\footnote{The questions  raised by this discussion have broader implications that go beyond conformal truncation itself.  If one can show that single-trace OPE data is sufficient to determine large $N$ RG flows in a given class of theories, then all bulk contact interactions in the AdS duals of these theories, such as $g_4$ (or e.g.~the $R^4$ term in gravity) are, in a certain sense, fully determined by the 3-pt interactions.  This has a natural interpretation in tree-level string theory, where one expects that knowledge of the 3-pt interactions for all string states determine the full string scattering amplitude.  But it appears very surprising from the point of view of AdS effective field theory, where contact interactions would seem to be independent parameters. }

\subsection{The Role of Zero Modes in a Holographic Model} 
\label{sec:MissingZero}

One can see intuitively that zero modes are the culprit behind the incorrect LC prediction above.  The problem arises from  the non-trivial background profile $\phi_{\rm cl}$, which is manifestly pure zero mode since it is momentum-independent.  
In this subsection, we will analyze  the nature of the missing zero mode contributions  in more detail. We  work in $d=2$ spacetime dimensions for simplicity.

We can bring the zero mode contributions back into view by starting with equal-time quantization and then taking the lightcone limit via an infinite boost.   For this purpose, we need only introduce a small momentum $q_- > 0$ for the relevant deformation:
\be
V = \lambda \int d x^- e^{i q_- x^-} \Ocal(x_-) + h .c. \equiv \lambda \Ocal(q_-).
\ee
The deformation now has time-like energy-momentum, which means we are performing equal-time quantization in some $q_-$-dependent frame.

We will see that in the limit $q_- \rightarrow 0$, the contributions from double-trace operators in the bulk model get pushed to infinitely high dimension, outside the space of the truncated basis used for  Hamiltonian truncation.  To start, we can write out the old-fashioned perturbation theory (OFPT) for the perturbation $V$ at second order:
\be
\Big[ \< \Ocal, p , \mu | H | \Ocal , p' , \mu' \> \Big]^{(2)} \sim \sum_\psi  \frac{\< \Ocal, p, \mu | V | \psi\> \< \psi | V | \Ocal, p', \mu' \> }{E_\psi - E_\Ocal} .
\label{eq:OFPT2nd}
\ee
The sum over $\psi$ is a sum over all states retained by the truncation. We will restrict our attention to the double-trace states, which are parameterized by their  twist $n$, spin $\ell$, momentum $P_-$, and invariant mass-squared $\mu^2$.  As shown in the previous subsection, the matrix elements of $V$ between a single-trace operator $\Ocal$ and a double-trace operator $[\Ocal^2]_{n,0}$ (with twist $n$ and, for simplicity,  spin 0) are proportional to a power $\nu = \Delta-1$ of the momentum $q_-$:
\be
 \< \Ocal, p , \mu | V | [\Ocal^2]_{n,0}, P, M \> \propto \delta(P_- - p_- - q_-)  q_-^\nu , \qquad \nu \equiv \Delta-1,
 \label{eq:BulkModelDoubleTrace}
 \ee
 and therefore vanish at $q_- = 0$ for $\Delta >1$.
 However, in addition to the sum over double-traces, there is also a divergent integral over their invariant mass-squared $M^2$. 
  The crucial point is that the infinite sum over {\it all} the double-traces resums into a function of $M^2$ that vanishes at large $M^2 \gg p_-/q_-$:
 \be
 \Big[ \< \Ocal, p , \mu | H | \Ocal , p' , \mu' \> \Big]^{(2)} \stackrel{M \gg \mu \atop p \gg q_-}{\sim} \int_0^\infty dM^2 \frac{q_- }{p_-} f(M^2 \frac{q_-}{p_-} ).
 \ee
  We relegate the explicit details of the sum and the function $f$ for the toy model to appendix \ref{app:OFPTdetails}, but the basic point is independent of its precise form.  
  
  Individual double-trace states and their descendants just contribute to a finite number of terms in the Taylor series of $f$; therefore they vanish at $q_- \rightarrow 0$, and their integral over $M^2$ diverges at finite $q_-$.  Both problems are solved in the infinite sum, where the integral over $M^2$ converges, and absorbs the $q_-$ dependence at small $q_-$.  In other words, there is a problem with the order of limits -- if we perform the infinite sum before taking $q_- \to 0$ then zero modes will be correctly included, but taking the lightcone limit before performing the sum discards all zero mode contributions.
  
  To summarize, at finite $q_-$, the ``zero mode'' contributions are present in the sum over physical double-trace states, but at $q_- \rightarrow 0$, their contributions are lifted from the spectrum. In the next section, we will introduce a prescription to recapture the zero mode contributions that get discarded by lightcone quantization, using only the correlators of the UV CFT fixed point.  A key lesson of the above analysis is that we will be trying to reintroduce contributions that, as $q_- \rightarrow 0$, no longer have any representation as a sum over physical intermediate states in the Hilbert space.


\section{Integrating Out the Zero Modes: A Prescription}
\label{sec:Prescription}

As compared to the standard, equal-time description, lightcone quantization can often provide striking simplifications.  But these advantages may come at a cost, because the lightcone appears oblivious to the complexities of vacuum structure, as it ignores zero modes and their mixing with other states.  Thus we need a prescription for detecting these zero mode contributions, determining if they affect observables, and including them where necessary.  In this section we will develop such a prescription and provide both a position and momentum space version.  Then in section \ref{sec:Applications} we will show how our prescription resolves a number of issues with lightcone quantization.

To motivate our prescription, we first study equal-time quantization in a frame with very large momentum $P_x$.  This will make it possible to see how the zero modes drop out as $P_x \to \infty$, but remain as additional $\delta(x^+)$ function contributions to correlators.  Our prescription identifies these delta functions and uses them to build a new lightcone Hamiltonian $H_\eff$ that includes the zero modes.  These delta functions can also be identified in momentum space as polynomial terms in the lightcone momenta.

\subsection{Argument for Prescription}

Consider a general CFT, which is then deformed by a relevant operator $\Ocal_R$.  A natural set of observables are the time-dependent two-point functions $\<\Ocal(t)\Ocal(0)\>$ of local UV CFT operators in the presence of this deformation.  We can construct these by inserting the unitary time evolution operator between two CFT basis states
\be
\<\Ocal,P_x,\mu|U(t,0)|\Ocal,P_x,\mu'\> \equiv \<\Ocal,P_x,\mu|\Tcal\{ e^{-i\int_0^t dt' [H_0 + V(t')]}\}|\Ocal,P_x,\mu'\>.
\ee
We can evaluate this expression by expanding $U(t,0)$ as the Dyson series
\be
U(t,0) = 1 - i \int_0^t dt_1 \, H(t_1) - \frac{1}{2} \int_0^t dt_1 dt_2 \, \Tcal \{ H(t_1) H(t_2) \} + \dots
\ee
By construction, the external basis states are eigenstates of the original CFT Hamiltonian $H_0$, which means we only need to consider the contributions of the deformation $V$ to this series,
\be
\begin{split}
\label{eq:DysonSeries}
\<\Ocal,P_x,\mu|U(t,0)|\Ocal,P_x,\mu'\> &\supset \<\Ocal,P_x,\mu|\Ocal,P_x,\mu'\> -i \int_0^t dt_1 \<\Ocal,P_x,\mu|V(t_1)|\Ocal,P_x,\mu'\> \\
& \qquad - \half \int_0^t dt_1 dt_2 \<\Ocal,P_x,\mu|\Tcal\{ V(t_1) V(t_2)\}|\Ocal,P_x,\mu'\> + \cdots
\end{split}
\ee
In order to compute any two-point function $\<\Ocal(t)\Ocal(0)\>$, in principle we need all correlation functions involving $n$ intermediate insertions of the deformation $\Ocal_R$.

For concreteness, let's focus specifically on the second-order term in this expansion, which corresponds to a  four-point function in the CFT. In equal-time quantization, we can compute this four-point function by inserting a complete set of intermediate states,
\be
\begin{split}
&\<\Ocal,P_x,\mu|V(t_1) V(t_2)|\Ocal,P_x,\mu'\> \\
& \qquad \qquad = \sum_{\psi} \int d\mu_\psi^2 \<\Ocal,P_x,\mu|V(t_1)|\psi,P_x,\mu_\psi\>\<\psi,P_x,\mu_\psi|V(t_2)|\Ocal,P_x,\mu'\>.
\end{split}
\label{eq:4ptDecomp}
\ee
The individual contributions of intermediate states correspond to momentum space three-point functions, or equivalently the matrix elements of the deformation.

However, if we take the limit $P_x \ra \infty$, we find that all intermediate contributions where $\De_\psi = \De + \De_R + n$ vanish. Roughly, such intermediate states, which correspond to lightcone zero modes, have matrix elements that behave like
\be
\< \Ocal, P_x, \mu | V(t_1) |\psi, P_x, \mu_\psi \> \sim \left( \frac{\mu^2 - \mu_\psi^2}{P_x} \right)^\alpha \qquad (P_x \ra \infty),
\ee
for some power $\alpha$. This is exactly what we saw for the bulk toy model in eq.~\eqref{eq:BulkModelDoubleTrace}, where the intermediate two-particle contributions all vanished in the lightcone limit. In that example, the small lightcone momentum transfer $q_-$ we introduced is equivalent to $\frac{\mu^2 - \mu_\psi^2}{P_x}$ in the large momentum limit,
\be
q_- = P_- - P_{\psi-} = \left( \sqrt{\mu^2+  P_x^2} - P_x \right) - \left( \sqrt{\mu_\psi^2+ P_x^2} - P_x \right) \stackrel{P_x \gg \mu}{\sim} \frac{\mu^2 - \mu_\psi^2}{P_x}.
\ee

The time-ordered two-point function $\<\Ocal(t) \Ocal(0)\>$ is independent of the choice of momentum frame, so these contributions must still be present in the full, physical result.  For any fixed $P_x$ these contributions are present in the sum over states, but as $P_x$ increases, their contributions come from larger and larger $\mu_\psi^2$, the invariant mass of the intermediate states.  At $P_x = \infty$, which is equivalent to working in lightcone quantization (see appendix~\ref{app:LCvsET}), the four-point function is \emph{no longer} reproduced as a sum over states.  In other words, in the lightcone limit $P_x \rightarrow \infty$, there are contributions to the four-point function that do not have a spectral function decomposition.  Momentum space three-point functions, which correspond to the naive set of Hamiltonian matrix elements, are \emph{not sufficient} to reproduce $\<\Ocal(t) \Ocal(0)\>$ in lightcone quantization.

The missing zero modes do not have a spectral representation in lightcone quantization.  Instead they appear as local terms in lightcone time, meaning that their contributions are proportional to $\delta(x^+)$.  We will see many explicit examples in section \ref{sec:Applications}.  When inserted in the LC version of the Dyson series in eq.~\eqref{eq:DysonSeries}, this delta function eliminates one of the integrals over time, reducing a second-order term in the Dyson series to the effective first-order  term
\be
\int_0^{x^+} dx_1^+ dx_2^+ \<\Ocal,P_-,\mu|\Tcal\{ V(x_1) V(x_2)\}|\Ocal,P_-,\mu'\> \sim \int_0^{x^+} dx_1^+ \<\Ocal,P_-,\mu|\de H_\eff(x_1)|\Ocal,P_-,\mu'\>.
\ee

We can reintroduce these missing contributions by \emph{defining} a new effective lightcone Hamiltonian via the derivative at $x^+=0$ of the unitary evolution operator $U(x^+,0)$:
\be
\boxed{H_\eff \equiv \lim_{x^+ \to 0} i \partial_+ U(x^+,0),}
\label{eq:prescription}
\ee
Matrix elements of the effective Hamiltonian $H_\eff$ are therefore written as a sum over $n$-point functions involving $n-2$ insertions of the relevant deformation $\Ocal_R$. When these higher-point functions are regular at $x^+=0$, only the term linear in $V$ contributes, because the region of integration $0 \le x_i^+ \le x^+$ shrinks to zero at $x^+=0$. In this case, our prescription reduces to the standard definition of Hamiltonian matrix elements in terms of three-point functions. However, when vanishing intermediate states lead to factors of $\de(x^+)$, the higher-point correlators can contribute even in the $x^+\rightarrow 0$ limit, modifying the naive lightcone Hamiltonian to include the effects of zero modes. Note that this prescription only relies on ``data'' derived from the correlation functions in the UV CFT.

 In order to see that the \emph{only} higher-order contributions to $H_\eff$ come from lightcone zero modes, let's look more explicitly at the four-point function contributions which \emph{do} have a spectral decomposition in lightcone quantization. These  can be rewritten in the form
\be
\begin{split}
&\<\Ocal,P_-,\mu|\Tcal\{V(x_1^+) V(x_2^+)\}|\Ocal,P_-,\mu\> \\
& \qquad \qquad \supset \sum_{\psi} \int_0^{\Lambda^2} d\bar{\mu}^2 \, \big| \<\Ocal,P_-,\mu | V | \psi, P_-, \mu_\psi\> \big|^2 \left( e^{i \frac{\bar{\mu}^2}{2P_-} x_{12}^+} \theta(x_{12}^+) +e^{-i \frac{\bar{\mu}^2}{2P_-} x_{12}^+} \theta(-x_{12}^+)  \right) \\
& \qquad \qquad = \int_0^{\Lambda^2} d\bar{\mu}^2 \, \rho(\mu,\bar{\mu}) \int dP_+ \, e^{i P_+ x_{12}^+} \left( \fr{i}{2 P_+ P_- - \bar{\mu}^2 + i\epsilon} + \fr{i}{2 P_+ P_- + \bar{\mu}^2 - i\epsilon} \right),
\end{split}
\label{eq:PrescriptionSF}
\ee
where we have taken both external states to have the same invariant mass $\mu$ for simplicity, we have defined $\bar{\mu}^2 \equiv \mu_\psi^2 - \mu^2$, and the four-point function spectral density is 
\be
\rho(\mu,\bar{\mu}) \equiv \sum_\psi \big|\< \Ocal,P_-,\mu | V | \psi,P_-,\mu_\psi \>\big|^2.
\ee
Note that in eq.~\eqref{eq:PrescriptionSF} we have written $\supset$ instead of $=$ because, crucially, not all contributions to this four-point function are contained in the sum over states in lightcone quantization.

The advantage of the last expression in eq.~\eqref{eq:PrescriptionSF} is that, for finite UV cutoff $\Lambda$, the integral over $\bar{\mu}$ is finite, since the integrand and the range of integration are finite, and therefore no $\delta(x_{12}^+)$ factors can be produced.  Manifestly, only terms in the four-point function that \emph{cannot} be written in this spectral function representation will contribute to our prescription for $H_\eff$.

While this discussion has been somewhat technical, at its core, our prescription can be understood as a simple matching procedure between equal-time and lightcone quantization. Specifically, the effective lightcone Hamiltonian $H_\eff$ is constructed such that that the resulting time-dependent two-point functions of local operators match those in equal-time quantization. In equal-time quantization, these two-point functions can be computed by inserting a complete set of states, which means that the three-point function contributions to the Hamiltonian are sufficient to reconstruct the resulting dynamics. However, in lightcone quantization, this is no longer true, precisely because we have \emph{removed some of the intermediate states from the Hilbert space} by discarding zero modes. Our definition for $H_\eff$ in terms of the unitary evolution operator is thus simply designed to add back in any contributions which were initially discarded in lightcone quantization.


\subsection{Momentum Space Diagnostic}
\label{sec:Plants}

Now that we have a general prescription for constructing the effective Hamiltonian from CFT correlators, we can develop a practical diagnostic for determining when this Hamiltonian receives higher-order contributions due to zero modes. Inserting our definition of $H_\eff$ from eq.~\eqref{eq:prescription} in between two basis states, we obtain the general matrix element expression
\be
\<\Ocal,P|H_\eff|\Ocal',P'\> \equiv \lim_{x^+\ra\infty}i  \p_+ \<\Ocal,P|U(x^+,0)|\Ocal',P'\>.
\ee
Consider a generic higher-point function appearing in the Dyson series expansion of the right-hand side
\be
\<\Ocal,P|U(x^+,0)|\Ocal',P'\> \supset \fr{(-i)^n}{n!} \int_0^{x^+} dx_1^+ \cdots dx_n^+ \<\Ocal,P|\Tcal\{V(x_1) \cdots V(x_n)\}|\Ocal',P'\>.
\ee
This correlator will only lead to a nonzero contribution to $H_\eff$ if it contains $n-1$ delta functions in $x_{ij}^+$. To see when these delta functions can arise, we can study the associated momentum space correlator
\be
\begin{split}
&\int d^dx_1 \cdots d^dx_n \, e^{i(q_1 \cdot x_1 + \cdots + q_n\cdot x_n)} \<\Ocal,P|\Tcal\{\Ocal_R(x_1) \cdots \Ocal_R(x_n)\}|\Ocal',P'\> \\
&\qquad \qquad \equiv (2\pi)^d \de^d\Big(P-P'+\sum_i q_i\Big) G(q_i,P,P').
\end{split}
\ee
Because the relevant deformation $\Ocal_R$ is a scalar, this function $G$ can only depend on the momenta $q_i$ via the Lorentz invariant combinations
\benn
q_i^2, \quad q_i\cdot q_j, \quad q_i \cdot P, \quad q_i \cdot P'.
\eenn
However, because the momenta $q_i$ correspond to insertions of the lightcone Hamiltonian, their ``spatial'' components $\vec{q}_i = (q_{i-}, \vec{q}_{i\perp})$ are all set to zero, eliminating the first two Lorentz invariant terms. The remaining two terms are actually equivalent due to conservation of lightcone momentum, so that the only nonzero Lorentz invariant combination for each $q_i$ is
\benn
q_i \cdot P = q_{i+} P_-.
\eenn
We therefore have the following simple test: we can only obtain $n-1$ delta functions (or derivatives of delta functions) in $x^+$ if the function $G$ contains a contribution which is \emph{analytic} in $q_i \cdot P$ (multiplied by an arbitrary function of $P, P'$).

\begin{figure}[t!]
\centering
\includegraphics[width=.32\linewidth]{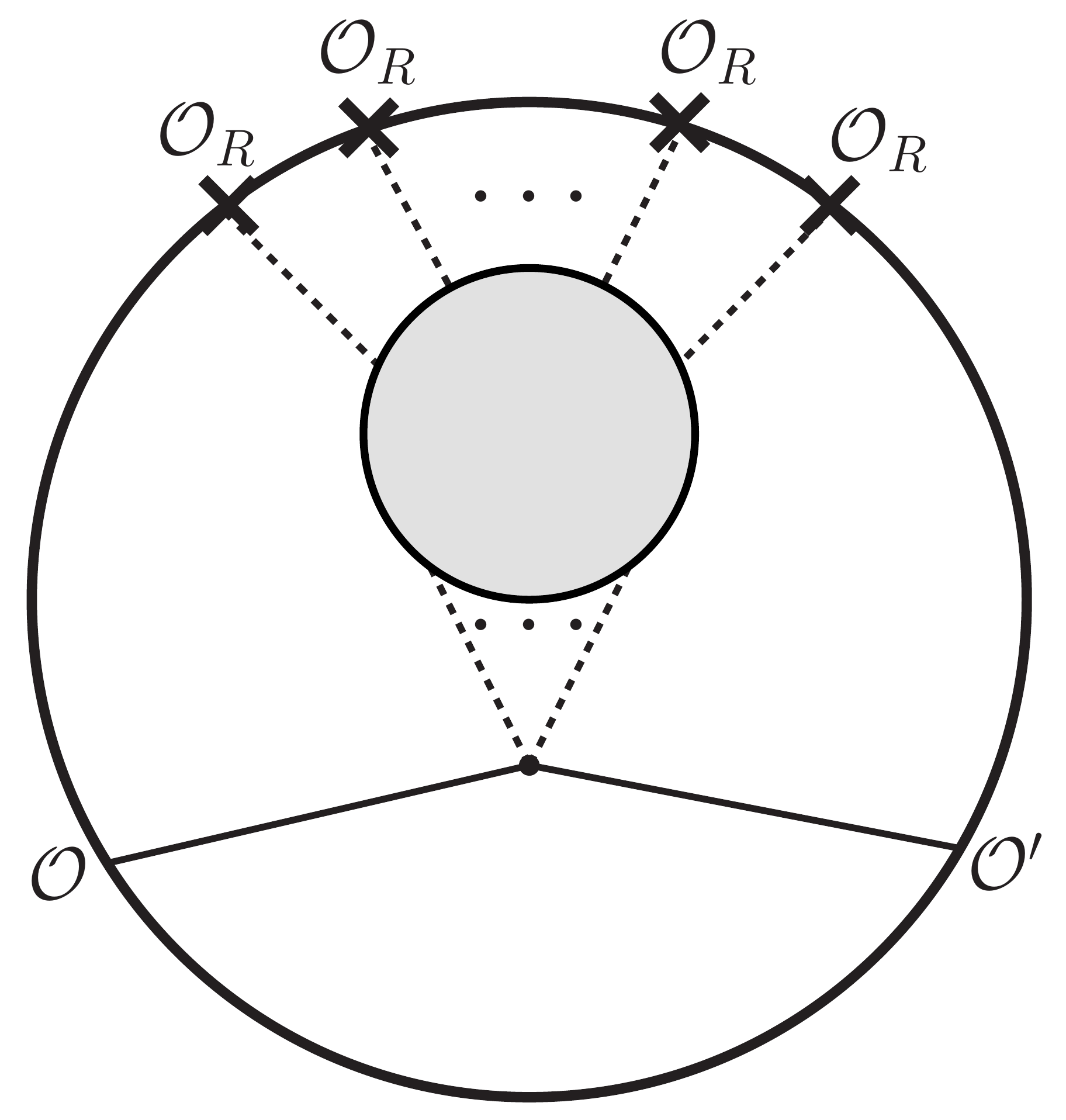}
\caption{General structure of ``plant'' diagrams which lead to effective Hamiltonian contributions in large $N$ theories. The zero modes (dashed lines) created by the relevant deformation $\Ocal_R$ must only connect to the physical states (solid lines) via a single contact interaction.}
\label{fig:PlantDiagram}
\end{figure}

For large $N$ theories, which have a weakly coupled bulk description, there is a particular class of Witten diagrams which automatically satisfy this test, and thus give rise to corrections to the effective Hamiltonian. Any ``plant'' diagram where the physical modes created by the external states connect to the zero modes created by $\Ocal_R$ through a single vertex (the ``base'' of the plant) corresponds to a polynomial in $q_i\cdot P$. As we can see from the schematic example in figure~\ref{fig:PlantDiagram}, these plant diagrams have a simple interpretation as the propagation of physical modes in the background bulk profile created by zero modes.

Similarly, if the UV CFT we're deforming is either free or has a weakly-coupled boundary description, such as a Banks-Zaks fixed point, then any Feynman diagram with this same plant structure will lead to corrections to the effective Hamiltonian. Unlike the large $N$ case, however, where the sum of plant diagrams create nonlocal interactions reproducing an entire bulk profile, the contributions from these boundary plant diagrams  correspond to local interactions in the field theory.
For example, a boundary diagram similar to figure~\ref{fig:PlantDiagram}, where only two physical propagators connect to the plant, simply gives rise to a mass counterterm for the physical modes. In other words, \emph{plant diagrams on the boundary only shift bare parameters in the Lagrangian}. In this case, the contributions from zero modes therefore don't affect any of the resulting dynamics,  as they only alter the map between UV parameters in the Lagrangian and the resulting IR
scales. Unless we are interested in this precise map, we can therefore safely ignore the contributions from boundary plant diagrams, to all orders in perturbation theory. As we discuss in section \ref{sec:phi4}, it is precisely this class of plant diagrams which explain observed discrepancies in perturbation theory for $\phi^4$ theory in equal-time and lightcone quantization.

More generally, we now have a straightforward diagnostic for determining which correlation functions (if any) contribute to the effective Hamiltonian, by looking at their momentum space dependence on $q_i$. For CFTs which are either perturbative or have a weakly-coupled AdS dual, this analysis becomes especially simple and can be performed at the level of diagrams.  In section \ref{sec:Applications} we will use our prescription to either resolve discrepancies between equal-time and lightcone quantization, or to demonstrate that such discrepancies are harmless.


\section{Examples and Applications}
\label{sec:Applications}


\subsection{Vacuum Energy}

We will now use our prescription for $H_\eff$ to study the vacuum energy for the deformation of a general CFT,
\be
\<H_\eff\> \equiv \lim_{x^+\ra0} i\p_+\<U(x^+)\>.
\ee
It will be easy to see that there are contributions to the vacuum energy from each order in the Dyson series expansion of the RHS. Consider the $n^{\mathrm{th}}$ order term
\begin{equation}
\Delta E_n
= \lim_{x^+\ra0} i\p_+ \int_0^{x^+} dx_1^+ \cdots dx_n^+ \<\Tcal\{V(x_1^+) \cdots V(x_n^+)\}\>.
\end{equation}
Following the analysis of section~\ref{sec:Plants}, we can determine whether this term contributes to the vacuum energy by studying the associated momentum space correlator $G(q_i)$,
\be
\<\Tcal\{V(x_1^+) \cdots V(x_n^+)\}\> = \lambda^n \int d q_{1+} \cdots d q_{n+} \, e^{i \sum_i q_{i+} x_i^{+}} iG(q_i) \, \delta^d\Big(\sum_i q_i\Big).
\ee
However, because all the Lorentz invariant scalar products $q_i \cdot q_j$ vanish, $G$ must be a constant,
\be
G(q_i) \equiv G_0,
\ee
resulting in a non-zero contribution to the vacuum energy,
\be
\Delta E_n = \lim_{x^+\ra0} i\p_+ \int_0^{x^+} dx_1^+ \cdots dx_n^+ \,i \lambda^n G_0 \prod_{i=1}^{n-1} \delta(x_i^+-x_n^+) \delta^{d-1}(0) = - \lambda^n G_0 \, \delta^{d-1}(0).
\ee

Our discussion has been very general, so we will now work out the contributions to the vacuum energy explicitly in the case of free field theory.  Some of the computations will also be useful as a warm-up for more complicated examples that we will study below.


\subsubsection*{Vacuum Energy in Free Field Theory}

Consider  a free scalar theory perturbed by a mass term $m^2 \phi^2$.  All diagrams that contribute to the vacuum matrix elements $\< {\rm vac} | H_{\rm eff}| {\rm vac}\>$ are built from one-loop diagrams with $\phi^2$ insertions included in the loop.  To evaluate their contribution to $H_{\rm eff}$, we will work in mixed position/momentum space,  keeping $x^+$ in position space and all other coordinates in momentum space. In mixed position/momentum space, each $\phi$ propagator can be written as
\be
G(x^+, \vec{k} ) = \int dk_+ \frac{e^{i k_+ x^+}}{ 2k_+ k_- - k_\perp^2 + i \epsilon}. 
\ee
Note that if $x^+ >0$, then we can close the contour in the upper half-plane, and the propagator vanishes unless $k_-<0$.  Similarly, if $x^+ < 0$, then we get zero unless $k_->0$.   The case where $x^+ = 0$ is more subtle and has to be treated carefully, and in fact our prescription dictates that all zero mode contributions come from this case.
 
A ``vacuum'' loop with $n$ mass terms has no external momenta flowing through the loop, so every propagator has the same $k_-$ and $\vec{k}_\perp$:
 \begin{equation}
\<\Tcal\{V(x_1^+) \cdots V(x_n^+)\}\> = m^{2n}   \delta^{d-1}(0) \int d k_- d^{d-2} k_\perp \CI , \quad  \CI \equiv \int \prod_{i=1}^n dk_{i+} \frac{i e^{ i k_{i+} (x_i^+- x_{i+1}^+)}}{2 k_{i+} k_- - k_{\perp}^2 + i \epsilon} .
 \end{equation}
Our argument above implies that  this contribution vanishes unless all $x_i^+$ coincide.  But when the $x_i^+$ coincide, we can obtain (formally infinite) delta function contributions.  We can integrate over the $x^+_i$ to calculate the coefficients of these $\delta(x_{i,i+1}^+)$ functions.  This integration forces all $k_{i+}$ to be identical, so we are simply left with
 \begin{equation}
 \int \left( \prod_{i=1}^{n-1} dx_i^+ \right) \CI = \int dk_+ \frac{(2\pi i)^n}{(2k_+ k_- - k_{\perp}^2 + i \epsilon)^n} .
 \label{eq:vacpres}
 \end{equation}
 This result is also a little subtle.  If $k_- \ne 0$, then there is an order-$n$ pole in one place, so we can close the contour on the other side of the real axis and see that we just get zero.  But if $k_-=0$, then the integration over $k_+$ diverges.  We have again identified a delta function, this time  $\delta(k_-)$.  We can integrate over $k_-$ to pick up the coefficient of the $\delta(k_-)$:
 \be
 \int dk_- \left( \prod_{i=1}^{n-1} dx_i^+ \right) \CI  = \int \frac{dk_+ dk_- (2\pi i)^n}{(2k_+ k_- - k_{\perp}^2 + i \epsilon)^n} 
   = (-2 \pi i)^{n+1} \int_0^\infty \frac{r dr}{(r^2 + k_{\perp}^2 )^n} 
    = \frac{(-2 \pi i)^{n+1} }{(n-1) k_{\perp}^{2(n-1)}} , \nn\\
\ee
where we have Wick rotated and changed to radial coordinates.  So finally we see that 
\be
\CI =  \frac{(-2 \pi i)^{n+1} }{(n-1) k_{\perp}^{2(n-1)}} \delta(k_-) \prod_{i=1}^{n-1} \delta(x_{i,i+1}^+) .
\ee
Therefore, these diagrams  contribute (only) to our prescription for $H_\eff$, and furthermore we see that their entire contribution comes from the $k_-=0$ modes. The full vacuum energy is the resummation of all possible such diagrams.  While we have focused on free field theories, this discussion would also apply to bubble diagrams in more general theories.

\subsection{Ising Model}

We have mainly focused on the case where zero modes are individual Fock space modes in an otherwise dynamical field. However, there are cases where an entire \emph{field} becomes non-dynamical in LC quantization and must be integrated out to generate an effective Hamiltonian for the remaining dynamical fields.

A simple illustrative example is the 2d Ising model. As is well known, a deformation of this CFT by the energy density $\eps$ is equivalent to free field theory of a massive fermion,
\be
\Lcal = \Lcal_{\textrm{Ising}} - m \eps = i\psi \p_+ \psi + i\chi \p_- \chi - \sqrt{2} m \chi\psi,
\label{eq:IsingL}
\ee
with the operator identification
\be
\eps = \sqrt{2} \chi \psi, \quad T \equiv T_{--} = i\psi \p_- \psi, \quad \overline{T} \equiv T_{++} = i\chi\p_+\chi.
\ee
From the Lagrangian, we see that in LC quantization the left-moving field $\chi$ has no kinetic term and is thus non-dynamical. Equivalently, its free equation of motion restricts $\chi$ to only be composed of zero modes,
\be
P_- \chi = 0.
\ee
We therefore need to integrate out $\chi$ to obtain an effective Lagrangian for the physical degrees of freedom built from $\psi$,
\be
\Lcal_\eff = i\psi \p_+ \psi - \fr{i}{2} m^2 \psi \fr{1}{\p_-} \psi.
\label{eq:FermEff}
\ee
As we'll now demonstrate, our prescription \emph{automatically} constructs this effective potential directly from the CFT correlation functions, without making any appeal to equations of motion.

From a conformal truncation perspective, the need to integrate out $\chi$ can first be seen when constructing the naive Hamiltonian from three-point functions. For example, if we look at the simplest matrix element, which corresponds to mixing between $\eps$ and $T$, we find
\be
\<\eps,P|V|T,P'\> = m \int d^2x_1 \, dx_2^- \, d^2x_3 \, e^{i(P\cdot x_1 - P'\cdot x_3)} \<\eps(x_1) \eps(x_2) T(x_3)\> = 0.
\ee
Just like in other free or large $N$ examples, this integral vanishes because the scaling dimensions of $\eps$ and $T$ are related by an integer,
\be
\De_T = 2\De_\eps.
\ee
This behavior continues for other three-point functions, such that we naively find that \emph{all} contributions to the Hamiltonian due to this deformation vanish in lightcone quantization.

However, using our prescription, we know that there may be corrections to the Hamiltonian from higher-point functions. For example, let's consider the matrix element between two insertions of the stress tensor component $T$, which is built solely from the dynamical field $\psi$. Using our prescription, the resulting matrix element can be written in terms of the unitary evolution operator,
\be
\<T,P|H_\eff|T,P'\> \equiv \lim_{x^+\ra0} i\p_+ \<T,P|U(x^+)|T,P'\>.
\ee
The insertion of the evolution operator can then be expanded into a Dyson series, turning this expression into a sum of correlation functions,
\be
\begin{split}
\<T,P|U(x^+)|T,P'\> &= \<T,P|T,P'\> - i \int_0^{x^+} dx_1^+ \<T,P|V(x_1)|T,P'\> \\
& \qquad - \, \fr{1}{2} \int_0^{x^+} dx_1^+ dx_2^+ \<T,P|\Tcal\{V(x_1) V(x_2)\}|T,P'\> + \cdots
\end{split}
\ee
For this particular example, the standard three-point function contribution to the Hamiltonian actually vanishes because the associated OPE coefficient is zero,
\be
\<T(x_1) \eps(x_2) T(x_3)\> = 0.
\ee
However, let's look more carefully at the next contribution in this series, due to the four-point function,
\be
\<T,P|\Tcal\{V(x_1) V(x_2)\}|T,P'\> = m^2 \int dx_1^- dx_2^- \<T,P|\Tcal\{\eps(x_1) \eps(x_2)\}|T,P'\>.
\ee
Because the external states are built only from $\psi$, we can use the fermion representation of $\eps$ to factorize this expression into two independent correlators,
\be
\<T,P|\Tcal\{\eps(x_1) \eps(x_2)\}|T,P'\> = -2\<\Tcal\{\chi(x_1)\chi(x_2)\}\> \cdot \<T,P|\Tcal\{\psi(x_1) \psi(x_2)\}|T,P'\>.
\ee
The time-ordered two-point function of $\chi$ in this expression is given by
\be
\<\Tcal\{\chi(x_1)\chi(x_2)\}\> = \fr{-i}{4\pi(x_{12}^+ - i\epsilon \, \sgn(x_{12}^-))} = \Pcal \left( \fr{-i}{4\pi x_{12}^+} \right) + \fr{1}{4} \de(x_{12}^+) \, \sgn(x_{12}^-),
\ee
where $\Pcal$ indicates the principal value. The propagator for $\chi$ thus gives rise to a delta function singularity in $x^+$, leading to a nonzero contribution to the effective Hamiltonian,
\be
\begin{split}
&\<T,P|\de H_\eff|T,P'\> = -\half \lim_{x^+\ra0} i\p_+ \int_0^{x^+} dx_1^+ dx_2^+ \<T,P|\Tcal\{V(x_1) V(x_2)\}|T,P'\> \\
& \quad = \fr{m^2}{4} \lim_{x^+\ra0} i\p_+ \int_0^{x^+} dx_1^+ dx_2^+ \de(x_{12}^+) \int dx_1^- dx_2^- \, \sgn(x_{12}^-) \<T,P|\Tcal\{\psi(x_1) \psi(x_2)\}|T,P'\> \\
& \quad = \fr{im^2}{4} \int dx_1^- dx_2^- \, \sgn(x_{12}^-) \<T,P|\Tcal\{\psi(x_1) \psi(x_2)\}|T,P'\>.
\end{split}
\ee

\begin{figure}[t!]
\centering
\includegraphics[width=.8\linewidth]{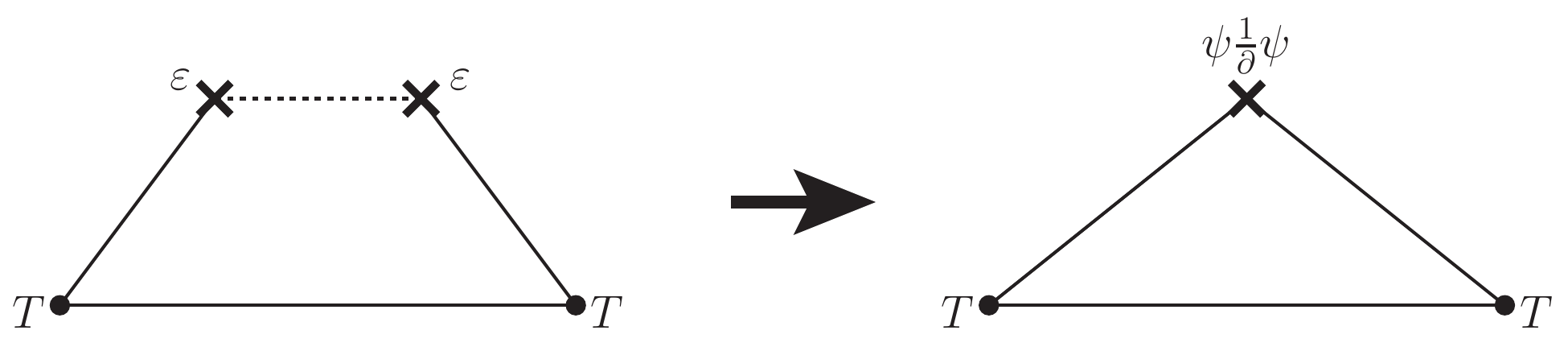}
\caption{Using our prescription, the Ising four-point function $\<T\eps\eps T\>$ gives rise to a Hamiltonian matrix element involving the effective interaction $\psi \fr{1}{\p} \psi$. This contribution arises due to the factor of $\de(x^+)$ in the $\chi$ propagator (dashed line).}
\label{fig:IsingExample}
\end{figure}

Not only do we obtain a nonvanishing contribution to the Hamiltonian, but the resulting matrix element is actually equivalent to that obtained by integrating out $\chi$,
\benn
\<T,P|\de H_\eff|T,P'\> = \fr{im^2}{2} \int dx^- \<T,P|\psi(x) \fr{1}{\p_-} \psi(x)|T,P'\>.
\eenn
Our prescription therefore automatically ``integrates out'' $\chi$ to reduce the original four-point function to an effective three-point function, as shown schematically in figure~\ref{fig:IsingExample}.

We can also understand the emergence of this delta function by looking at the associated momentum space structure of this four-point function, which takes the schematic form
\be
G(q,P,P') = \left(\fr{q_+}{q_+ q_- - i\epsilon}\right) f(q_-,P,P').
\ee
The only dependence on the lightcone energy $q_+$ associated with the Hamiltonian insertions comes from the $\chi$ propagator, which factors out from the rest of the correlator. The real part of this propagator is manifestly independent of $q_+$, leading to a delta function in $x^+$.

If we repeat this analysis for the higher-point functions in the Dyson series, we find that there are no other contributions to $H_\eff$, which can easily be seen by dimensional analysis. Each pair of $\eps$ insertions brings two lightcone time integrals but only one delta function, so at higher orders the number of delta function singularities is insufficient to overcome the suppression as $x^+ \ra 0$. We therefore only need to consider four-point functions to obtain the full effective Hamiltonian for this particular theory.


\subsection{$\phi^4$ Theory}
\label{sec:phi4}

Let's now apply our prescription to a simple example where the resulting IR theory is interacting. Our UV CFT is simply free field theory involving a single massless scalar field, which we then deform by adding a mass term and a quartic interaction,
\be
V = \int d^{d-1}x \left( \half m^2 \phi^2 + \fr{1}{4!} \lambda \phi^4 \right).
\label{eq:Phi4Potential}
\ee
This particular example will allow us to demonstrate how our prescription resolves multiple known problems in lightcone quantization: the discrepancy between bare couplings in ET and LC identified by Burkardt in~\cite{Burkardt:1992sz}, the divergent contributions due to zero modes discussed by Hellerman and Polchinski in~\cite{Hellerman:1997yu}, and the insensitivity to tadpoles.

\subsubsection*{Matching Bare Parameters in LC and ET}

\begin{figure}[t!]
\centering
\includegraphics[width=.35\linewidth]{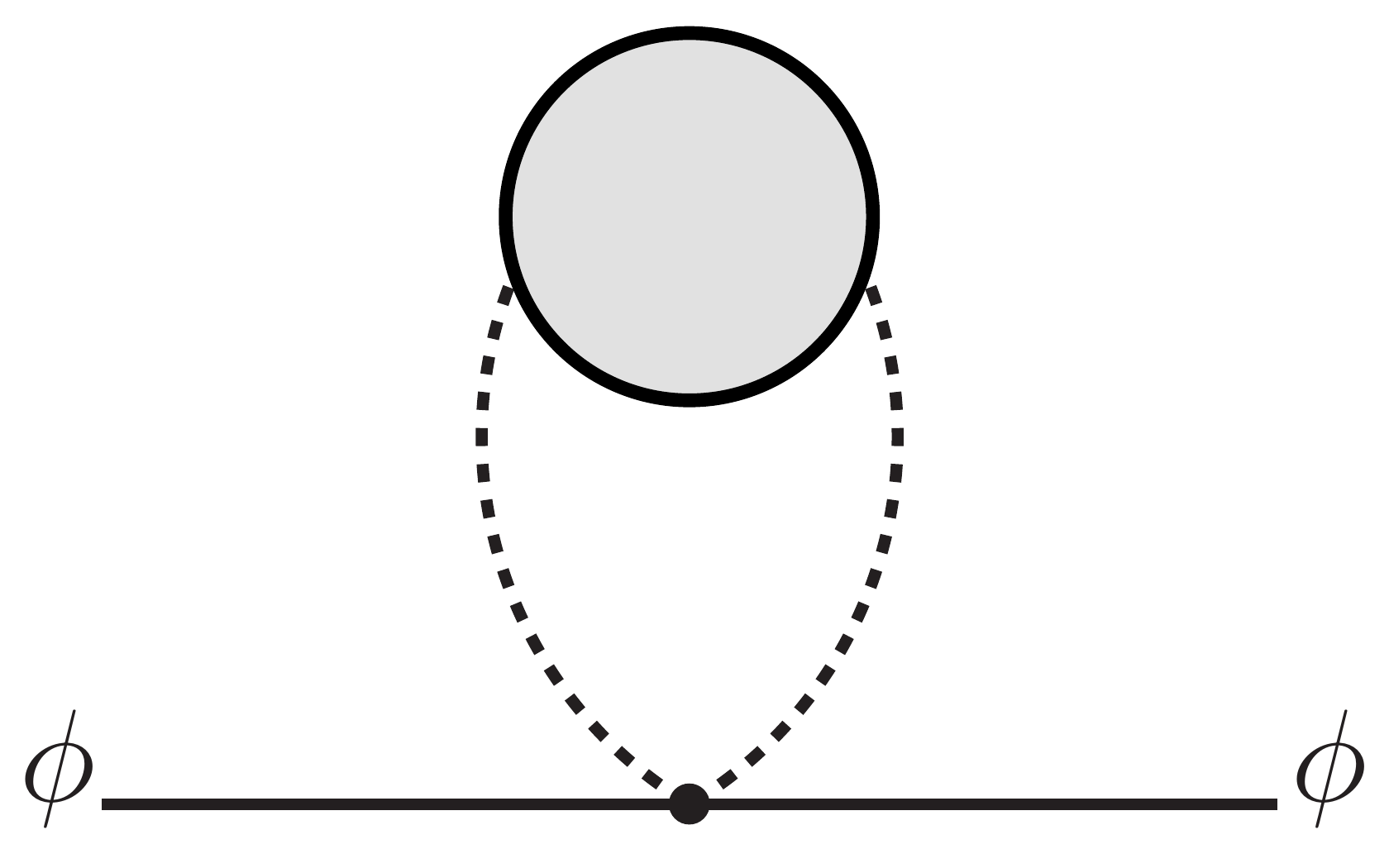}
\caption{General structure of plant diagrams which contribute to the effective Hamiltonian in $\phi^4$ theory. These higher-point correlation functions correspond to Feynman diagrams which contribute to the one-particle mass in equal-time quantization but vanish in lightcone quantization. By including these terms in $H_\eff$, our prescription eliminates the naive discrepancy between the two quantization schemes.}
\label{fig:Burkardt}
\end{figure}

As discussed in section~\ref{sec:BareCoupling}, there is a disagreement in the resulting spectrum of $\phi^4$ theory if we use the same bare parameters in ET and LC quantization. However, at least in perturbation theory, this discrepancy can be removed by shifting the LC bare mass by a counterterm proportional to the expectation value of $\phi^2$ in the interacting theory~\cite{Burkardt:1992sz},
\be
m^2_\LC = m^2_\ET + \frac{\lambda}{2} \<\phi^2\>.
\label{eq:counterterm}
\ee
This discrepancy between ET and LC quantization only arises because we are missing the effects of zero modes on the Hamiltonian. In other words, the naive expression for $V_\LC$, with the original bare parameters, is incomplete and needs the additional corrections from higher-point functions, which naturally lead to the shift in eq.~\eqref{eq:counterterm}. We can reproduce the result (\ref{eq:counterterm}) by applying our prescription to the one-particle matrix element,
\benn
\<\phi,P|H_\eff|\phi,P'\> = \lim_{x^+\ra0} i\p_+ \sum_n \fr{(-i)^n}{n!} \int_0^{x^+} dx_1^+ \cdots dx_n^+ \<\phi,P|\Tcal\{V(x_1) \cdots V(x_n)\}|\phi,P'\>.
\eenn
If we look at the various terms in this sum, we find that there is a class of contributions, shown schematically in figure~\ref{fig:Burkardt}, which all have the ``plant diagram'' structure discussed in section~\ref{sec:Plants}. These higher-point correlators therefore contain delta function singularities and result in additional contributions to the LC Hamiltonian. More importantly, these correlation functions precisely correspond to the Feynman diagrams used in \cite{Burkardt:1992sz} to determine the counterterm in eq.~(\ref{eq:counterterm}).  So in this case, our prescription just reduces to the result of this earlier work. 

We have tested (\ref{eq:counterterm}) explicitly up to $\CO(\lambda^5)$ for the case $d=2$ by numerically computing the mass gap in LC and in ET quantization. First, we discuss the perturbative results.  The LC mass gap $\mu_{\rm gap}$ is 
\be
&&\mu^2_{\rm gap, LC} = m_{\rm LC}^2\left[ 1 + \sum_{n=2}^\infty c_n \left( \frac{\lambda}{m_{\rm LC}^2} \right)^n\right] ,
\nn\\
&&  c_2 = -\frac{3}{2} , \ c_3 = \frac{9}{\pi}, \ c_4 = -11.5198, \ c_5 = 53.62,
\ee
whereas the ET mass gap is\footnote{The ET results were obtained by a combination of numerical results from Hamiltonian truncation and from explicit computations of the Feynman diagrams. The latter were obtained by private communication from  M.~Serone and G.~Spada, whom we thank for sharing their preliminary results.} 
\be
&& \mu^2_{\rm gap, ET} = m^2_{\rm ET} \left[ 1+ \sum_{n=2}^\infty \tilde{c}_n \left( \frac{\lambda}{m_{\rm ET}^2} \right)^n\right] ,
\nn\\
&&  \tilde{c}_2 = -\frac{3}{2} , \ \tilde{c}_3 = \frac{9}{\pi}+ \frac{63 \zeta(3)}{2\pi^3}, \  \tilde{c}_4 = -14.656, \ \tilde{c}_5 = 65.97.
\ee
Finally, the vev $\<\phi^2\>$ is
\be
& & \< \phi^2 \>  = \frac{63 \zeta(3)}{ \pi^3} \lambda^2  -  \frac{513 \zeta(3)}{ \pi^4} \lambda^3+ 15.2612 \lambda^4 + \CO(\lambda^5). 
\ee
\begin{figure}[t!]
\begin{center}
\includegraphics[width=0.45\textwidth]{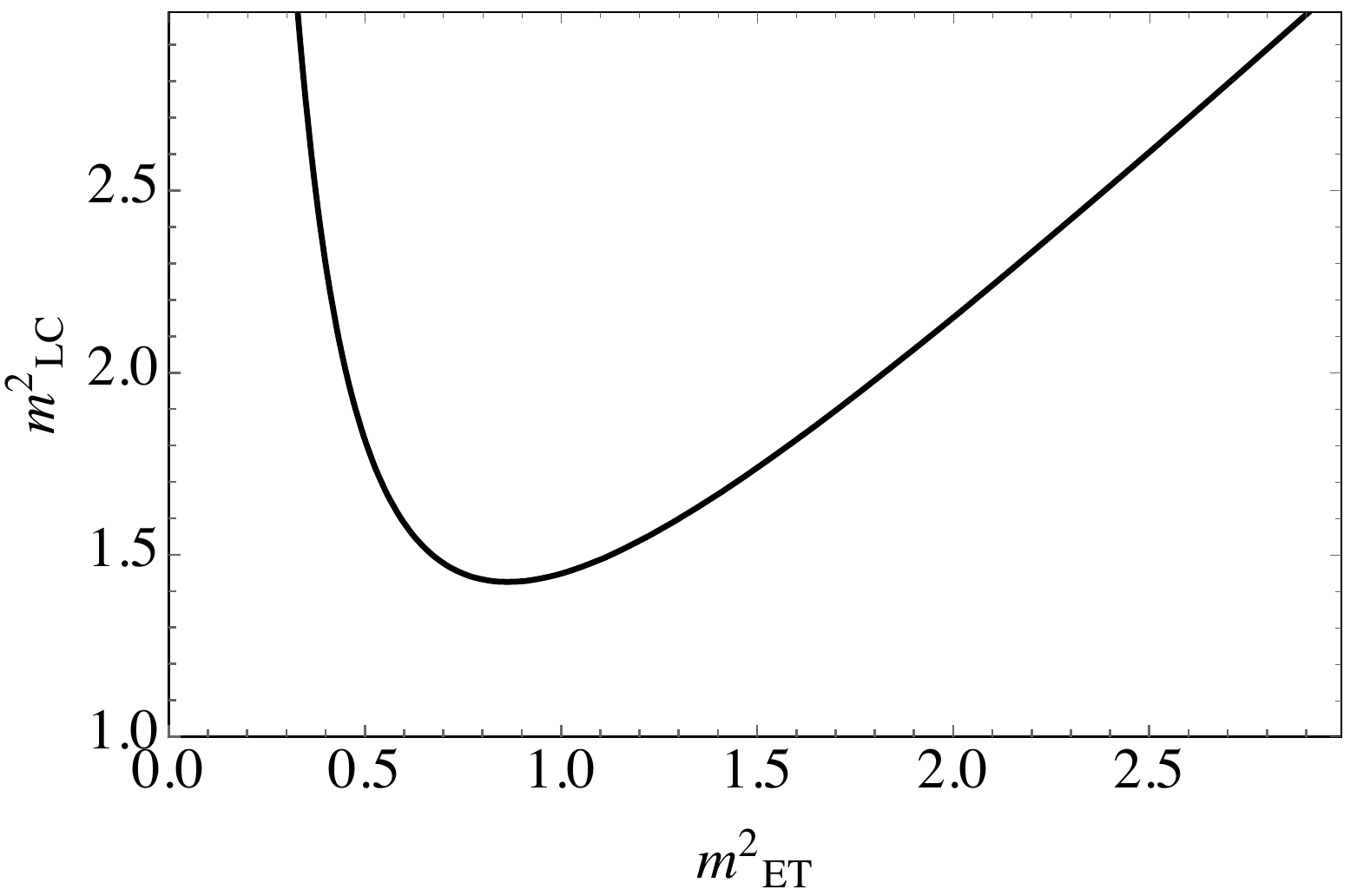}
\includegraphics[width=0.45\textwidth]{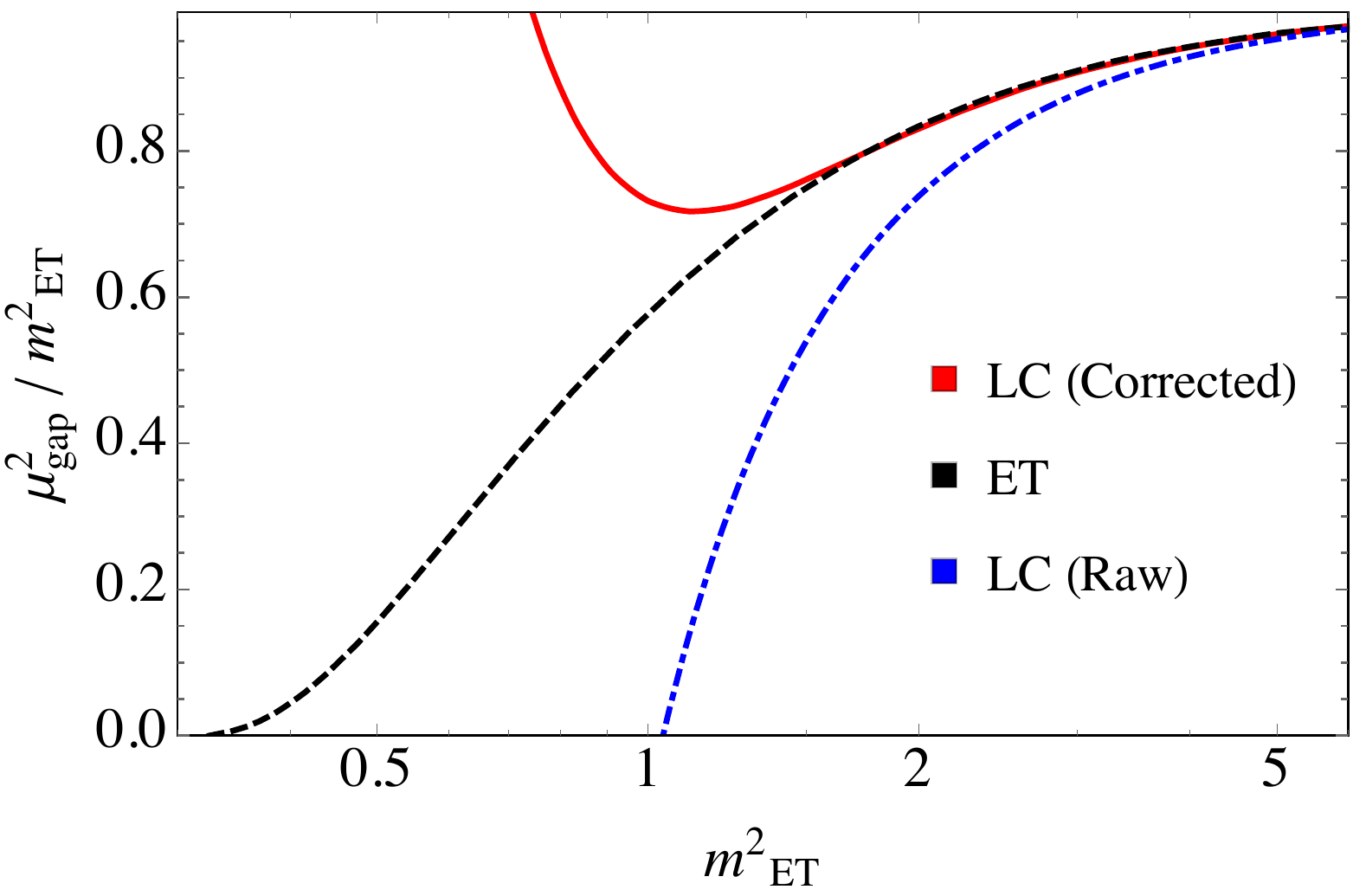}
\end{center}
\caption{{\it Left:} The predicted value for $m_{\rm LC}^2$ from (\ref{eq:counterterm}) as a function of $m_{\rm ET}^2$, with $\lambda=1$ fixed.  At large $m_{\rm ET}^2$ and $m_{\rm LC}^2$, the relation is well-described by perturbation theory (in $\tfr{\lambda}{m^2}$) and is well-behaved. However, at small $m_{\rm ET}^2$, the function is not monotonic, and as a result there are some real values of $m_{\rm LC}^2$ that correspond to two different real $m_{\rm ET}^2$  values, and some that do not correspond to {\it any} real $m_{\rm ET}^2$ values, indicating that the prescription cannot be correct in this regime. {\it Right:} The predicted mass gap $\mu_{\rm gap}^2$ in units of the bare mass, for the ET computation (black, dashed), the raw LC computation (blue, dot-dashed), and for the LC computation  corrected by (\ref{eq:counterterm}) (red, solid).   The corrected LC result does significantly better than the raw result at large $m_{\rm ET}^2$ (small $\frac{\lambda}{m_{\rm ET}^2}$), but starts to turn back upwards at small $m_{\rm ET}^2$ at the same point that $m_{\rm LC}^2(m_{\rm ET}^2)$ (left plot) does, and disagrees completely for smaller $m^2_\ET$.  }
\label{fig:msqLCvsET}
\end{figure}
Substituting (\ref{eq:counterterm}) into the LC expression reproduces the ET coefficients analytically up to $\CO(\lambda^3)$, and to within $0.2\% [4.4\%]$ at $\CO(\lambda^4) [\CO(\lambda^5)]$:
\be
c_2 \rightarrow -\frac{3}{2}, \ c_3 \rightarrow \frac{9}{\pi} + \frac{63 \zeta(3)}{2 \pi^3}, \ c_4 \rightarrow -14.685, \ c_5 \rightarrow 63.08.
\ee

While this perturbative check of \eqref{eq:counterterm} is an encouraging sign, we have nevertheless found experimentally that it appears to fail at the non-perturbative level.  The most serious issue is that as one decreases the bare ET mass-squared $m_{\rm ET}^2$ with $\lambda$ fixed,\footnote{The bare parameter $\lambda$ is easily matched between ET and LC since the theory is super-renormalizable and $\lambda$ is just the leading small 2-to-2 scattering amplitude at high energies in both quantizations.} the vev $\< \phi^2\>$ increases, and eventually the `matching' $m_{\rm LC}^2$ value defined via~\eqref{eq:counterterm} turns around and starts to {\it increase} for \emph{decreasing} $m_{\rm ET}^2$, as shown in figure~\ref{fig:msqLCvsET}.  Therefore, if the prescription were exactly correct, then by inspection we could choose two different values of $m_{\rm ET}^2$ that correspond to the same value $m_{\rm LC}^2$, which means that both $m_{\rm ET}^2$ values would have to have the \emph{same} physical predictions.  However, no such redundancy is seen in the numerical ET analysis, and this matching procedure thus fails for values of $m^2_\ET$ beyond the turnaround point. We leave a more detailed analysis of the interpretation and consequences of this result for future work.

\subsubsection*{No Hellerman-Polchinski Corrections}

In \cite{Hellerman:1997yu} Hellerman and Polchinski pointed out additional possible zero mode contributions, beyond those captured by figure~\ref{fig:Burkardt}. In that work, they were specifically interested in studying the effects of zero modes in the framework of discrete lightcone quantization (DLCQ) \cite{Pauli:1985pv,Pauli:1985ps}, in which the ``spatial'' direction $x^-$ is compactified. They demonstrated that, in this DLCQ framework, internal loops involving zero modes appeared to generically lead to IR divergences, precisely due to the $\de(x^+)$ structure of zero mode propagators.

\begin{figure}[t!]
\centering
\includegraphics[width=.9\linewidth]{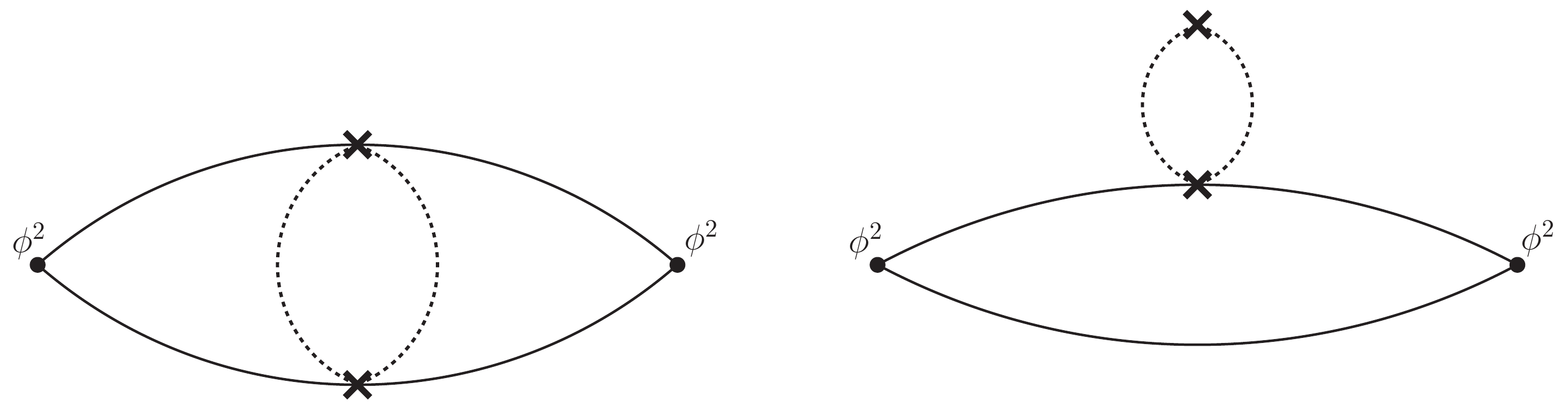}
\caption{Two possible corrections to the effective lightcone Hamiltonian due to four-point functions. By inserting a complete set of states, we find that the spectral decomposition of the left diagram survives in the infinite momentum limit, such that there are no analytic terms in $q_{i+}$ and thus no contributions to the effective Hamiltonian. The spectral decomposition of the right diagram vanishes in this limit, giving rise to delta function singularities which correct the lightcone Hamiltonian.}
\label{fig:HP}
\end{figure}

While we are instead interested in studying conformal truncation at infinite volume, one might worry that such divergences give rise to additional contributions to the effective lightcone Hamiltonian, beyond those that can be accounted for with a shift in the bare mass. For example, consider the diagram on the left in figure~\ref{fig:HP}, where two physical modes (solid lines) exchange a loop of zero modes (dashed lines). From the perspective of our new prescription, this diagram corresponds to a four-point function contribution in the Dyson series expansion,
\be
\<\phi^2,P|U(x^+)|\phi^2,P'\> \supset -\half \int_0^{x^+} dx_1^+ dx_2^+ \<\phi^2,P|\Tcal\{V(x_1) V(x_2)\}|\phi^2,P'\>.
\ee

However, if we repeat the analysis of section~\ref{sec:Plants} and study the momentum space form of this particular correlation function, we find that the loop integral gives rise to non-analytic dependence on $q_i \cdot P$, where $q_i$ are the momenta of the $V$ insertions. When we Fourier transform with respect to $q_{i+}$, we therefore do not obtain delta functions in $x^+$.

We can understand this behavior by inserting a complete set of intermediate states into this four-point function,
\be
\<\phi^2,P|V(x_1) V(x_2)|\phi^2,P'\> = \sum_\Ocal \int \fr{d^dP_\Ocal}{(2\pi)^d} \<\phi^2,P|V(x_1)|\Ocal,P_\Ocal\> \<\Ocal,P_\Ocal|V(x_2)|\phi^2,P'\>.
\ee
In ET quantization, this particular diagram only receives contributions from four-particle intermediate states. If we then take the infinite momentum limit, we find that these intermediate contributions do \emph{not} vanish. This correlation function thus retains its spectral decomposition in LC quantization and does not give rise to any terms which are analytic in $q_{i+}$. Phrased more simply, this process is already captured by the three-point function contributions to the lightcone Hamiltonian, with no additional four-point function contribution needed.

We can contrast this example with the diagram on the right in figure~\ref{fig:HP}, which is simply a two-particle generalization of the mass shift diagrams in figure~\ref{fig:Burkardt}. If we look at the spectral decomposition of this four-point function in ET quantization, we find that it also only receives contributions from four-particle intermediate states. However, if we again take the infinite momentum limit, we find that \emph{all} of these contributions vanish, because the associated operator scaling dimensions are related to that of $\phi^2$ by an integer,
\be
\De_\Ocal = 2\De_{\phi^2} + n \quad \Rightarrow \quad \lim_{|P_x|\ra\infty} \<\phi^2,P|\phi^2|\Ocal,P_\Ocal\> = 0.
\ee
This lack of a spectral decomposition in LC quantization leads to terms which are analytic in $q_{i+}$, which in turn leads to a correction to the effective Hamiltonian.

More generally, we find that the only correlation functions which contribute to the effective Hamiltonian are the mass renormalization diagrams in figure~\ref{fig:Burkardt} (and their higher-particle generalizations).

\subsubsection*{Effects from Tadpoles}

Finally, we can deform our theory by a tadpole term,
\be
\de V = \int d^{d-1}x \, g\phi.
\ee
From a Lagrangian perspective, it's obvious that we simply need to shift our field $\phi$ by the acquired expectation value $v$ in order to move to the minimum of the new potential. This shift then generates a cubic term and corrects the bare mass,
\be
g\phi + \half m^2 \phi^2 + \fr{1}{4!} \lambda \phi^4 \, \underset{\phi\ra\phi+v}{\Rightarrow} \, \half \left( m^2 + \half \lambda v^2 \right) \phi^2 + \fr{1}{3!} \lambda v \phi^3 + \fr{1}{4!} \lambda \phi^4
\ee
While the naive LC Hamiltonian built from three-point functions completely misses this effect (since all matrix elements involving the $\phi$ deformation vanish), it is straightforward to see that the higher-point contributions in $H_\eff$ automatically generate these shifts in bare parameters.

\begin{figure}[t!]
\centering
\includegraphics[width=.9\linewidth]{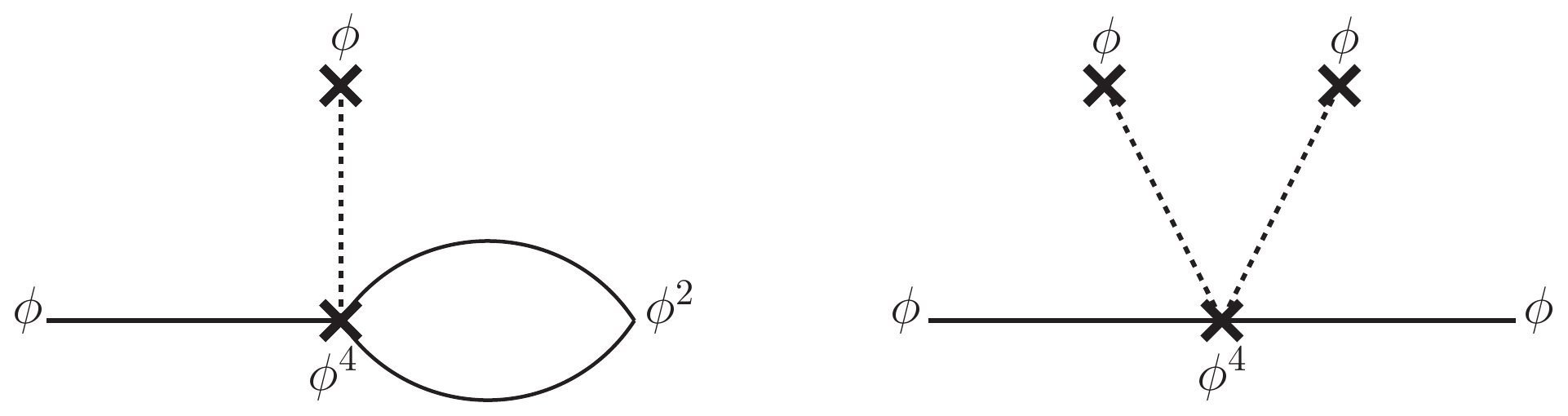}
\caption{Leading corrections to the cubic interaction (left) and bare mass (right) in $H_\eff$ due to a tadpole term. Including all diagrams of these two types in the LC Hamiltonian is equivalent to performing the field redefinition $\phi \ra \phi+v$.}
\label{fig:Tadpoles}
\end{figure}

For example, consider the matrix element contribution from the four-point function
\be
\<\phi,P|\Tcal\{V(x_1) V(x_2)\}|\phi^2,P'\> \supset g \lambda \int d^{d-1}x_1 \, d^{d-1} x_2 \<\phi,P|\Tcal\{\phi(x_1) \phi^4(x_2)\}|\phi^2,P'\>,
\ee
which corresponds to the left diagram in figure~\ref{fig:Tadpoles}. This diagram has the familiar plant diagram structure, which means it contains a delta function associated with the zero mode propagator (dashed line) and thus contributes to $H_\eff$. As we can see, the resulting matrix element clearly corresponds to a cubic interaction, mixing the one- and two-particle states.

Similarly, we can obtain the leading contribution to the mass shift by looking at the five-point function
\be
\begin{split}
&\<\phi,P|\Tcal\{V(x_1) V(x_2) V(x_3)\}|\phi,P'\> \\
& \qquad \qquad \supset g^2 \lambda \int d^{d-1}x_1 \, d^{d-1} x_2 \, d^{d-1}x_3 \<\phi,P|\Tcal\{\phi(x_1) \phi(x_2) \phi^4(x_3)\}|\phi,P'\>,
\end{split}
\ee
which corresponds to the right diagram in figure~\ref{fig:Tadpoles}. The two zero mode propagators each contain a $\de(x^+)$ function factor, leading to a non-zero correction to the bare mass.

Of course, to obtain the full $H_\eff$ we need to include the infinite set of higher-point functions that fall into these two classes. However, we can already see diagrammatically that performing this sum is equivalent to computing the VEV of $\phi$. Our prescription therefore automatically ``redefines'' $\phi$ to account for the presence of a tadpole.


\subsection{Holographic Models}

We now turn to the case of theories at large $N$, in order to demonstrate how our prescription resolves the problem encountered in section~\ref{sec:Problem}. Specifically, let's again consider the example of a large $N$ CFT$_d$ dual to a $\phi^4$ effective theory in AdS$_{d+1}$,
\be
\Lcal_{\textrm{bulk}} = \half (\p\phi)^2 - \half m^2 \phi^2 - \fr{1}{4} \fr{g_4}{N^2} \phi^4.
\ee
As discussed in section~\ref{sec:Problem}, if we deform this theory by the single-trace operator $\Ocal$ dual to the bulk field $\phi$,
\be
\label{eq:lbdy}
\Lcal_{\textrm{bdy}} = \Lcal_\CFT - \lambda N \Ocal,
\ee
we naively find that all contributions to the lightcone Hamiltonian vanish as $N \ra \infty$.

\begin{figure}[t!]
\centering
\includegraphics[width=.3\linewidth]{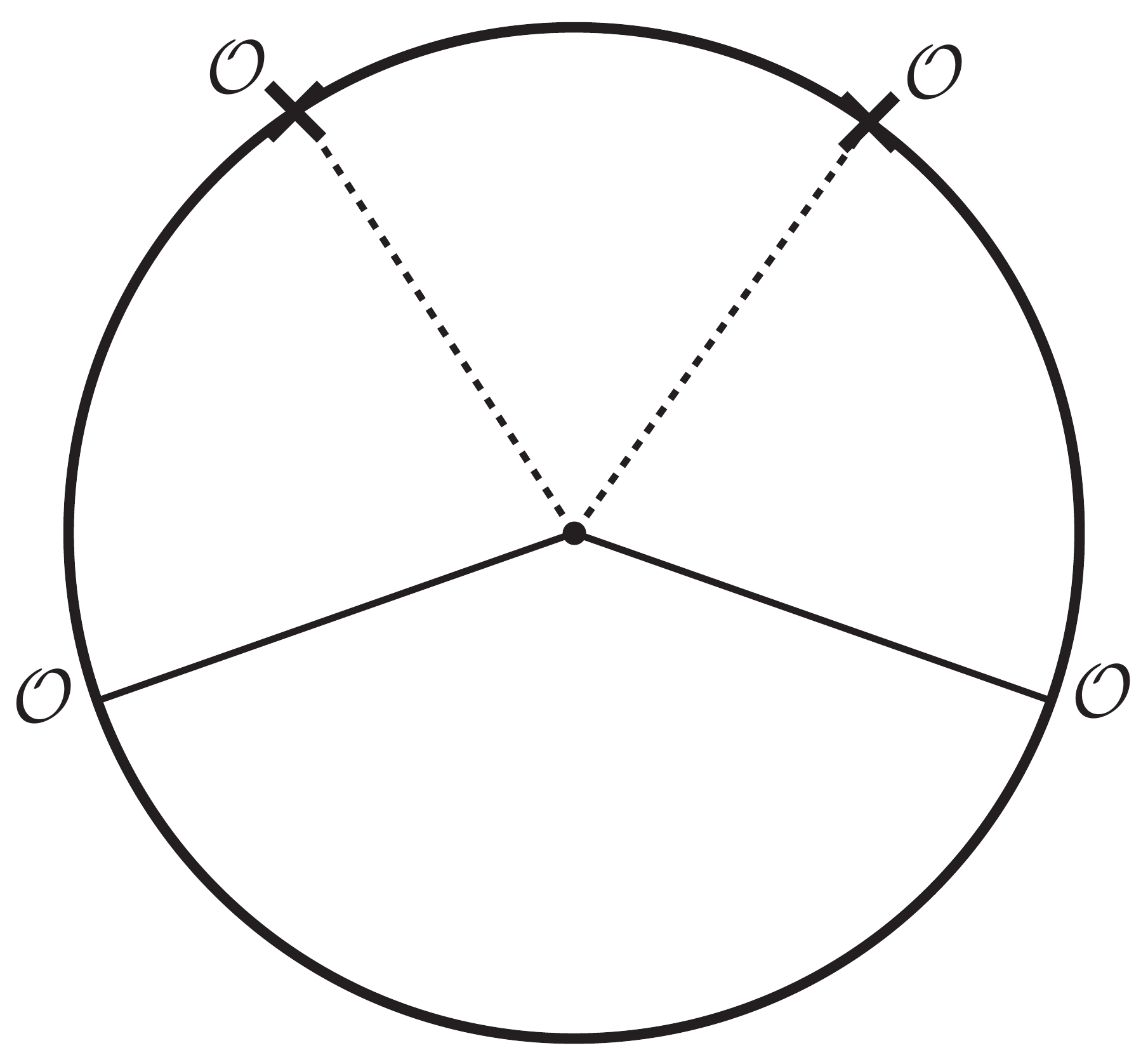}
\caption{Witten diagram for the leading effective Hamiltonian contribution due to a bulk quartic interaction. The bulk-to-boundary propagators of the two zero modes (dashed lines) have delta-function singularities in $x^+$, leading to a nonvanishing Hamiltonian matrix element for the physical external states (solid lines).}
\label{fig:BulkQuartic}
\end{figure}

However, let's now use our  prescription to construct the effective Hamiltonian, just as in prior examples. Focusing on the matrix element between two insertions of the single-trace operator $\Ocal$, we again have the possible four-point function contribution
\be
\<\Ocal,P|\de H_\eff|\Ocal,P'\> = - \fr{1}{2} \lim_{x^+\ra0} i\p_+ \int_0^{x^+} dx_1^+ dx_2^+ \<\Ocal,P|\Tcal\{V(x_1) V(x_2)\}|\Ocal,P'\>.
\label{eq:HeffQuartic}
\ee
We can compute the underlying position-space four-point function via AdS perturbation theory, where the leading correction due to the bulk interaction corresponds to the tree-level Witten diagram shown in figure~\ref{fig:BulkQuartic}.

As we can see, this diagram clearly has the ``plant'' structure discussed in section~\ref{sec:Plants}, which means its momentum space expression is analytic in the lightcone energy $q_{i+}$ associated with the two zero modes. In fact, because this particular bulk process is a simple contact interaction, the resulting function has \emph{no} dependence on $q_{i+}$. This four-point function therefore contains a delta function in $x^+$ and provides a nonzero contribution to the effective Hamiltonian.

We can also see the emergence of this delta function explicitly by directly computing the zero mode bulk-to-boundary propagators. So long as the single-trace scaling dimension $\De > \fr{d}{2}$, there are no IR divergences when integrating over the spatial directions, resulting in the simple expression
\be
K_\De(\vec{q}=0,x^+,z) = \int d^{d-1} x \left( \fr{z}{x^2 - z^2 - i\epsilon} \right)^\De = z^{d-\De} \, \de(x^+),
\ee
where we've suppressed any overall coefficients. We thus see that, as expected, propagators of bulk zero modes have delta-function singularities in $x^+$.

Our prescription therefore allows us to construct the effective Hamiltonian for a single-trace deformation even though the naive three-point function contributions vanish. Note that even though this resulting matrix element doesn't mix single-trace operators with multi-trace ones, it still relies on the multi-trace OPE coefficients contained in the four-point function. Our prescription thus confirms that for this particular toy example, our analysis of large $N$ theories in section~\ref{sec:LCAdvantages} was too naive. If there are delta-function singularities in higher-point functions of single-trace operators, then lightcone conformal truncation requires the full set of planar limit OPE coefficients, just as in equal-time quantization.

\begin{figure}[t!]
\centering
\includegraphics[width=.95\linewidth]{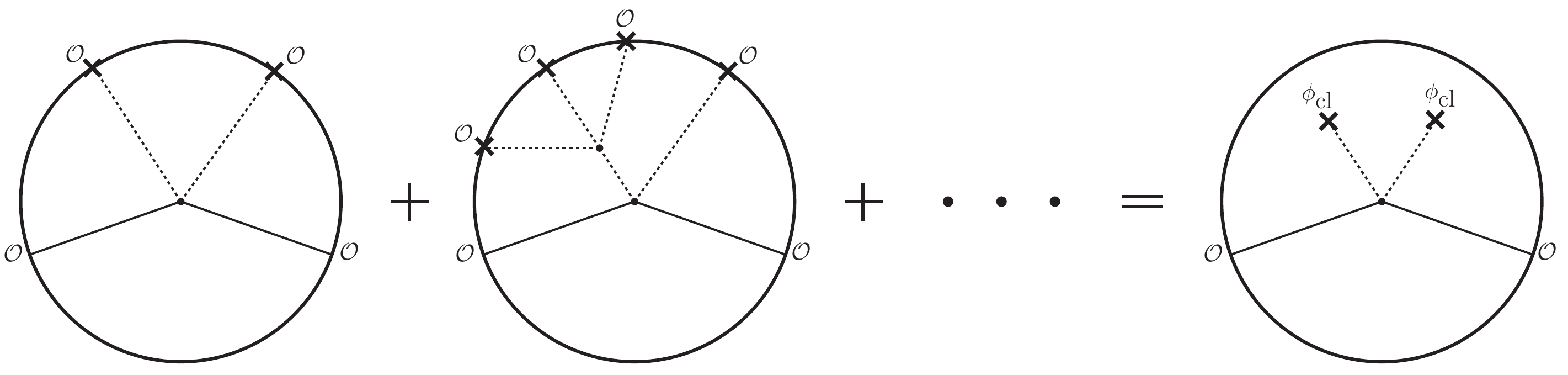}
\caption{The higher-point function contributions to $H_\eff$ corresponding to AdS ``plant diagrams'' resum to reproduce the bulk profile $\phi_{\textrm{cl}}$ which solves the bulk equations of motion.}
\label{fig:EOMfromPlants}
\end{figure}

The effective Hamiltonian for this bulk theory doesn't only receive a contribution from the four-point function, but actually has an infinite number of contributions coming from additional higher-point functions, all of which are necessary to correctly capture the IR physics. Let's analyze these higher-point corrections more carefully, in order to determine the structure of the resulting $H_\eff$. Consider the $n^{\mathrm{th}}$ order term in the Dyson series expansion,
\be
\<\Ocal,P|H_\eff|\Ocal,P'\> \supset \frac{(-i)^n }{n!} \lim_{x^+\ra0} i \partial_+ \int_0^{x^+} dx^+_1 \dots dx^+_n   
\braket{\Ocal, P}{\Tcal\{V(x_1) \dots V(x_n)\}}{\Ocal, P'}.
\label{eq:dysonP}
\ee
At leading order in the large $N$ limit, this term potentially receives Witten diagram contributions with the same plant structure as the four-point function, where the external states are connected to a single bulk vertex, from which a ``tree'' of zero modes grows towards the boundary to connect to the Hamiltonian insertions, as shown in figure~\ref{fig:EOMfromPlants}. 

All diagrams of this type are manifestly analytic in $q_{i+}$, which means they all provide a nonzero contribution to the effective Hamiltonian. The resummation of this infinite set of diagrams is actually equivalent to solving the equation of motion for $\phi$ in the bulk. This was anticipated in section 
\ref{sec:Problem}, where we demonstrated that ``naive'' conformal truncation does not capture the IR physics arising from a deformation of the bulk profile. In other words, resumming all of the higher-point function contributions to the Dyson series results in the full effective Hamiltonian
\be
\<\Ocal,P|H_\eff|\Ocal,P'\> = N^2 \de^{d-1}(P-P') \int \frac{d z}{z^{d+1}} \, S_{\rm bulk}''[\phi_{\rm cl}(z)] \, 
K_\De(P,z) K_\De(P',z),
\label{eq:deltah}
\ee
where $\phi_{\rm cl}$ is the solution of the bulk equation of motion with a boundary source,
\begin{equation}
\label{eq:phicl}
S'_{\rm bulk}[\phi_{\rm cl}(z)] = 0  \,, \qquad  \phi_{\rm cl}(z) \stackrel{z\sim 0}{\sim}  \lambda N z^{d-\Delta} 
+ \alpha z^{\Delta},
\end{equation}
and $K_\De$ is the momentum space bulk-to-boundary propagator for the physical modes created by the external states,
\be
K_\De(P,z) = \int d^d x \, e^{-iP\cdot x} \left( \fr{z}{x^2 - z^2 - i\epsilon \, \sgn (x^+)} \right)^\De = \mu^{\De-\fr{d}{2}} z^{\fr{d}{2}} J_{\De-\fr{d}{2}}(\mu z).
\ee
Diagonalizing $H_\eff$ therefore amounts to finding the spectrum of perturbations around the saddle point 
$\phi_{\rm cl}$ in the semiclassical large $N$ limit. This makes it clear that an infinite number of terms in the Dyson series are actually necessary to compute $H_\eff$ exactly, as the series expansion corresponds to an expansion of $\phi_{\rm cl}$ in powers of the boundary source $\lambda$, shown schematically in figure~\ref{fig:EOMfromPlants}. 

With knowledge of the bulk action, one may be able to reconstruct $\phi_{\rm cl}$ and compute $H_{\rm eff}$ directly. 
In appendix \ref{app:SUSY}, we discuss a simple supersymmetric model where this can be done analytically. 
However, in more physically realistic cases one might just
have access to the CFT correlators in eq.~\eqref{eq:dysonP}. In that case, one might hope that truncating the Dyson series 
to a fixed order in $\lambda$ could be sufficient to compute the Hamiltonian to a reasonable accuracy. 
However, there are two obstructions. First, the resummed series may not be analytic in $\lambda$. In a
bulk model, this manifests itself with the presence of a boundary VEV, corresponding to the $\alpha$ term in 
eq.~\eqref{eq:phicl}. Second, the individual series expansion terms in eq.~\eqref{eq:dysonP} can be IR divergent. 
For example, in any large $N$ model with local bulk interactions deformed by a single trace boundary deformation, 
the $n^{\mathrm{th}}$ order term in the Dyson series is proportional to
\begin{equation}
\label{eq:rthorder}
\<\Ocal,P|\de H_\eff^{(n)}|\Ocal,P'\> \sim \int \frac{d z}{z^{d+1}} z^{n(d - \Delta)} z^d J_{\De-\fr{d}{2}}(\mu z) J_{\De-\fr{d}{2}}(\mu' z),
\end{equation}
which diverges for $n > \fr{2}{d-\De}$. On the other hand, we know that the exact expression in eq.~\eqref{eq:deltah} is convergent,
as $\phi_{\rm cl}$ approaches a constant in the IR. Therefore, resumming the whole series is necessary to obtain a 
finite Hamiltonian.

Alternatively, one can introduce an IR regulator. One possibility is to introduce an IR brane at $z = z_{\textrm{IR}}$,
so that the integrals \eqref{eq:rthorder} will be finite by construction.
In that case, we can derive a rough estimate of the highest order term in the Dyson series
\begin{equation}
n_{\max} \gtrsim \lambda \Lambda_{\rm IR}^{\Delta - d}
\end{equation}
needed to access QFT observables (e.g. spectral densities) down to the IR scale $\Lambda_{\rm IR} \sim z_{\rm IR}^{-1}$.


\subsection{$O(N)$ Model}
\label{sec:ON}

As a final example of how to apply the prescription for $H_{\rm eff}$, in this section we consider a simple deformation of the $O(N)$ model at  large $N$. The $O(N)$ CFT can be defined via the explicit action \cite{Moshe:2003xn}
\begin{equation}
S = \int d^d x \left[ \frac{1}{2} \left(\partial_\mu \phi_i \right)^2 - \frac{1}{2} r \phi_i^2 - \frac{1}{N} \frac{u}{4!} \left(\phi_i^2 \right)^2 
\right] \,,  
\label{eq:ONLag}
\end{equation}
where $i =1 , \cdots, N$ and we tune to the critical point $r=0$. There is a free fixed point, $u=0$, as well as an interacting one with $u \ne 0$.  We will first focus  on the free fixed point; at the end of this section, we will describe the generalization to the interacting case.  We will deform by the singlet operator $\phi^2$, so that
\begin{equation} 
S = S_{\rm CFT} + \lambda N \int d^d x \, \phi^2 ,
\label{}
\end{equation}
and in what follows we drop the $i$ index on $\phi$ for simplicity.
We will see  that in this model, no contributions to the effective Hamiltonian arise, i.e. $H_{\rm eff}=H$, except for a renormalization of the vacuum energy.

\subsubsection*{Four-point Function} 
 
We begin by considering the contribution to $H_{\eff}$ at second order in the Dyson series for some external operators $\CO, \CO'$ that carry non-vanishing total momentum $p_-$.  We can try to extract the $\delta(x^+)$ coefficient from the corresponding four-point function by integrating over a small window around where the mass deformations coincide.  The four-point function in mixed position/momentum space contains two insertions, one at $x_1^+$ and one at $x_2^+$.  The $x_2^+$ dependence appears in two propagators, which together are
\be
\CI \equiv \int dk_+ dk_+' \frac{e^{i x_{12}^+ k_+ + i x_{23}^+ k'_+}}{(2k_+ k_- - k_\perp^2 + i \epsilon)(2k_+' k_- - k_\perp^2 + i \epsilon)}
\ee
We integrate over $x_2^+$ over a small window around $x_1^+$:
\be
\CI' \equiv \int_{x_1^+-\delta}^{x_1^+ + \delta} dx_2^+ \CI = e^{i k_+ ' x_{13}^+} \int dk_+ dk_+' \frac{ e^{ i (k_+ - k_+') \delta} - e^{- i (k_+ - k_+') \delta}}{i(k_+ - k_+' + i \epsilon)(2k_+ k_- - k_\perp^2 + i \epsilon)(2k_+' k_- - k_\perp^2 + i \epsilon)} . \nn\\
\ee
Next, we do the $k_+$ integral. We can take the $k_-$ to be non-negative, so both poles are below the real axis.  Therefore, only the term with the $- i k_+\delta$ in the exponent contributes:
\be
\CI' = 2\pi\int dk_+' \frac{ \left(  e^{i (k_+ '  - \frac{k_\perp^2}{2k_-}) \delta}e^{- \frac{\delta \epsilon}{2k_-}}-1 \right) }{2k_+' k_- - k_\perp^2 + i \epsilon} \frac{e^{i k_+ ' x_{13}^+}}{(2k_+' k_- - k_\perp^2 + i \epsilon)}
\ee
Now, the key question is what happens when we take $\delta \rightarrow 0$: do we get zero or not?  Note that if we assume $k_- >0$, then the answer vanishes at $\delta \rightarrow 0$:
\be
\lim_{\delta \rightarrow 0} \CI' = 0, \qquad (k_- > 0).
\ee
However, if $k_- =0$, then the $e^{- \frac{ \delta \epsilon}{2k_-}}$ causes the first term to shut down, and we instead find
\be
\lim_{\delta \rightarrow 0} \CI' = \int dk_+' \frac{2 \pi}{k_\perp^4}e^{i k_+ ' x_{13}^+}, \qquad (k_- = 0). 
\ee
Therefore, we explicitly find that the contributions to $H_{\rm eff}$ from the four-point function vanish when there is non-zero momentum $k_-$ flowing through the propagators between the relevant operators.       

\begin{figure}[t!]
\begin{center}
\includegraphics[width=0.65\textwidth]{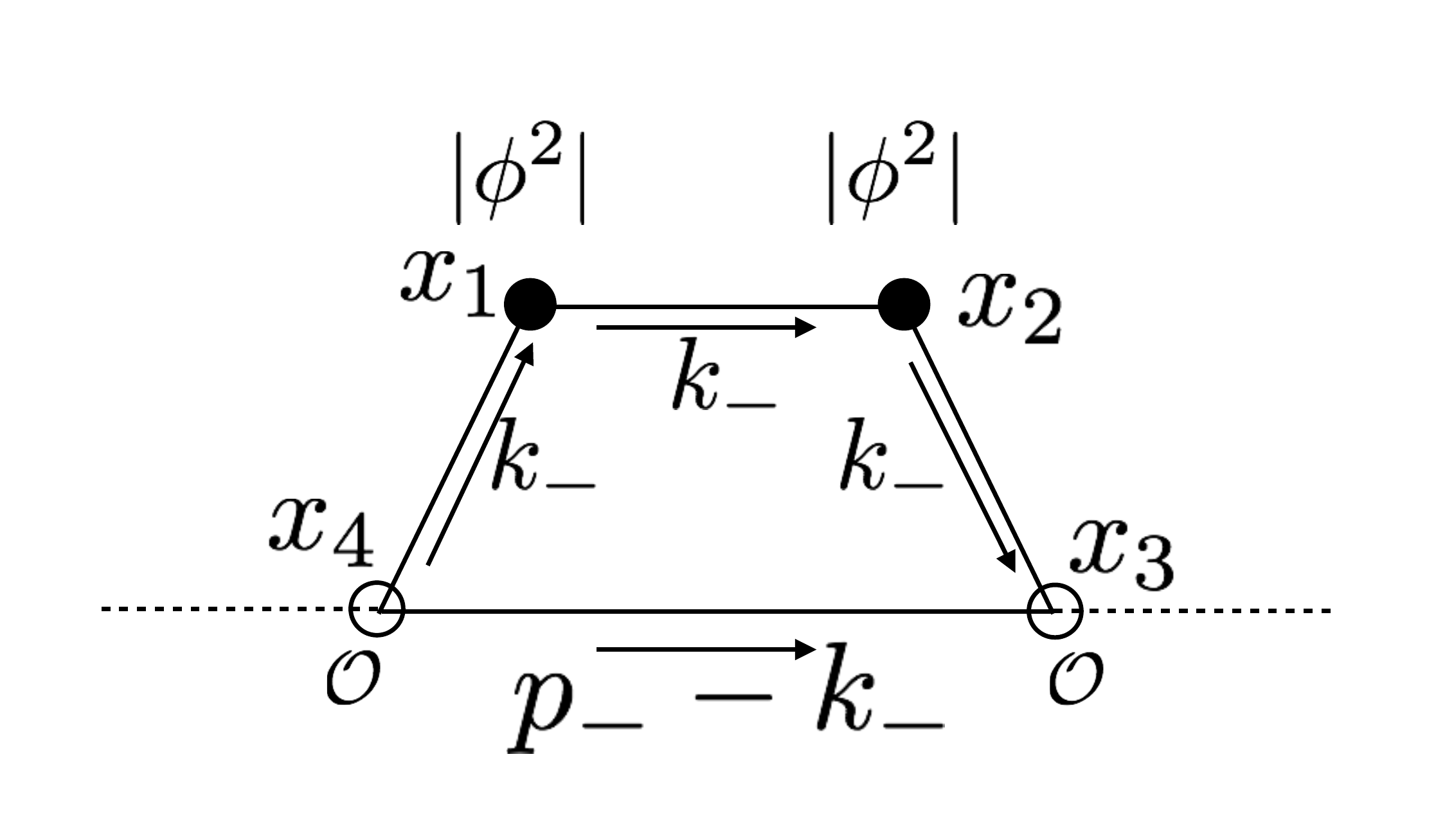}
\caption{Four-point diagram in mixed position/momentum space, contributing to $H_{\rm eff}$ for external states $\CO$ at second order in the Dyson series in the free $O(N)$ model with a mass-squared $\phi^2$ deformation.}
\label{fig:trapezoid}
\end{center}
\end{figure}

This result still allows for the possibility that a contribution arises that is purely localized at $k_-=0$.  Since $k_-=0$ is a point of measure zero in the integral over $k_-$, the only way it can contribute a finite amount is by having a $\delta(k_-)$ function localized contribution in the $dk_- $ integrand.  To establish that no such $\delta(k_-)$ contributions are present, we can set $k_-=0$ inside the integrand and show that we get a finite result.  So, let us again start with the four-point function with some general two-particle external states $\CO$ and $\CO'$.\footnote{At infinite $N$, the correlators of higher-particle states, e.g. $(\phi_i \phi_i)^2$, factor into the correlators of the two-particle states.}  Without loss of generality, for any tensor operator $\CO_{\mu_1, \dots, \mu_\ell}$, we can always choose the component with all minuses, i.e. $\CO_{-  \dots -}$.  In the two-particle operator case, such operators are linear combinations of $\partial_-$ derivatives acting on $\phi_i \phi_i$ (e.g. $(\partial_-^2 \phi_i) \phi_i + (\partial_- \phi_i)^2$), and in the one-loop diagram this linear combination simply introduces extra dependence on $k_-$ given by the corresponding polynomial $f_\CO(k_-)$.\footnote{In the free theory, or the interacting theory at infinite $N$, operators like $(\partial_\mu \phi)^2$ with internal contracted indices can always be reduced using the equations of motion to a linear combination of primaries and descendants of primaries that have only $\partial_-$ derivatives.} Consider now setting $k_-=0$, and following the propagators around the closed loop, as depicting in figure~\ref{fig:trapezoid}.  The first propagator, from the external operator at $x_4^+$ to the mass insertion at $x_1^+$, just becomes a $\delta(x_{41}^+$):
\be
\int dk_+ \frac{e^{i k_+ x_{41}^+}}{2k_+ k_- - k_\perp^2 + i \epsilon} \stackrel{k_-=0}{\rightarrow} \delta(x_{41}^+).
\ee
Similarly, the propagator between the two mass insertions just becomes $\delta(x_{12}^+)$, and the propagator between the second insertion and the final state becomes $\delta(x_{23}^+)$.  The last propagator, between the two external states, is however qualitatively different, since it has some external $p_-$ flowing through it:
\begin{equation}
f_\CO(k_-)  f_{\CO'}(k_-) (k_-)\int dk_+ \frac{e^{i k_+ x_{34}^+}}{2k_+ (k_- - p_-) - k_\perp^2 + i \epsilon} \stackrel{k_-=0}{\rightarrow}f_\CO(0)  f_{\CO'}(0) \int dk_+ \frac{1}{-2k_+ p_- - k_\perp^2+ i\epsilon}.
\end{equation}
We have included the polynomials $f_\CO(k_-), f_{\CO'}(k_-)$ corresponding to the external states, but since these are finite they can be ignored.  In addition, we have set the factor in the numerator to 1, since the other propagators became a product of $\delta$ functions $\propto \delta(x_{41}^+) \delta(x_{12}^+) \delta(x_{23}^+) = \delta(x_{41}^+) \delta(x_{12}^+) \delta(x_{43}^+)$, and so set $x_{34}^+=0$.  The integral on the RHS above is logarithmically divergent at large $k_+$.  Physically, a logarithmic divergence is not strong enough just based on dimensional analysis to correspond to a $\delta(k_-)$ singularity evaluated at $k_-=0$. To be more mathematically precise, we can note that the result is finite if we choose a two-sided regulator $-\Lambda < k_+ < \Lambda'$ with finite $\Lambda/\Lambda'$ in the limit $\Lambda \rightarrow 0$. Equivalently, if we choose a Gaussian regulator, $\sim e^{- \delta^2 k_+^2}$, then the RHS has a finite $\delta \rightarrow 0$ limit of $- \frac{i \pi}{2 |p_-|}$. 
 Since this result is finite at $k_-=0$, there is no $\delta(k_-)$ factor.\footnote{Note that this logic relied crucially on the presence of the $p_-$ in the last propagator.  In the vacuum bubble, $p_-=0$, and the last propagator at $k_-$ just produces another $\delta(x_{34}^+)$.  Since there is already a $\delta(x_{34}^+)$ from the chain of the other propagators, the final result contains $\delta(x_{34}^+)\delta(x_{34}^+)=\delta(x_{34}^+)\delta(0)$, which does have the required strength singularity to indicate the presence of a $\delta(k_-)$ term.  As a final comment, one may worry about divergences from the $k_\perp$ integral. However, one can always (and may be forced to) introduce a regulator on $k_\perp$. Since this regulator is boost invariant, it cannot introduce additional $k_-$ or $x_+$ dependence, and so cannot introduce either $\delta(k_-)$ or $\delta(x_+)$ functions.}

\subsubsection*{Higher-point Functions}

Finally, we argue that there are no contributions to $H_{\rm eff}$ from higher-point functions (other than the vacuum energy), by reducing higher point functions to the same form as the four-point function.

 First, since mass insertions $\phi^2$ have two legs, we can think of one ``coming in'' and one ``going out'' and we can follow them around the diagram in a chain until they eventually end by connecting to the external state.  Consider chains that are at least three insertions long.  We are only interested in $\delta(x^+)$ function contributions, so consider the case where all the times of the operators coincide.  If we take any insertion in the middle of the chain, it has $k_-$ flowing into and out of it, so it cannot have both its legs pointing into the past or into the future - if it does, then the diagram vanishes.  This means that we can integrate over its time, because the only possible contributions come from when it lines up with the time of the other operators in the tower.  But this leads to a significant simplification, if we look at the two propagator factors coming out of it:
\be
\int dx^+_i \int dk_+ dk'_+ \frac{e^{i k_+ (x_{i-1}^+ - x_{i}^+) +i k_+' (x_i^+ - x_{i+1}^+)}}{(2k_+ k_- - k_\perp^2+i \epsilon)(2k_+' k_- - k_\perp^2+i \epsilon)}  = \int dk_+  \frac{e^{i k_+ (x_{i-1}^+ - x_{i+1}^+) }}{(2k_+ k_- - k_\perp^2+i \epsilon)^2} . 
\ee 
The RHS of the above expression is the same form as a single propagator, except that the denominator is squared.  Clearly, repeating this procedure for $n$ such internal insertions just gives
\be
\int dk_+  \frac{e^{i k_+ (x_{i}^+ - x_{i+n}^+) }}{(2k_+ k_- - k_\perp^2+i \epsilon)^n} 
\label{eq:nprops} 
\ee
Therefore, all the propagators between all the mass terms just collapse to a single modified propagator of the above form.  The result is equivalent to a four-point function with this modified propagator between the two mass terms, and at $k_-=0$ the modified propagator simply has extra powers of $k_\perp$ compared to the original propagator.  The argument in the previous subsection then immediately applies.

\subsubsection*{Interacting Case}

Here, we will briefly describe how the above arguments can be generalized to the interacting fixed point of (\ref{eq:ONLag}). As usual, is it useful to rewrite the action in terms of an auxiliary field $\sigma$:
\be
S \cong \int d^d x \left[ \frac{1}{2} \left(\partial_\mu \phi_i \right)^2 - \frac{1}{2} r' \phi_i^2 - \frac{1}{2}\sigma \phi_i^2 + \frac{N}{u} \sigma^2 
\right] \,.  
\ee

At infinite $N$, the $\sigma$ two-point function can be computed in closed form.  At the critical point, it can be summarized by the fact that $\sigma$ becomes a primary dimension $\Delta = 2$ operator, so $\<\sigma(q) \sigma(-q)\> \propto q$. Planar 1PI correlators of the relevant deformation $\phi^2$ are just one-loop diagrams with $\phi^2$ insertions.  At a diagrammatic level, the main difference between a theory of $N$ free bosons and the interacting $\CO(N)$ CFT is that in the interacting theory there are $\sigma$ propagators attached to each $\phi^2$ insertion.  In momentum space, these are trivial to include. 

The prescription (\ref{eq:prescription}), on the other hand, expresses the correlators as functions of lightcone time.  Since contributions to $H_{\rm eff}$ arise only from $\delta(x^+)$ function dependence in the correlators, we can isolate these contributions either by working in momentum space and considering the limit where the energy $q_+$ of the relevant operator insertions goes to infinity, or we can work directly in real space and integrate over an infinitesimal window in time.  We will mostly use the latter strategy.  In this case, there is a useful simplification that arises when we restrict our attention to such $\delta(x^+)$ contributions, that allows us to take correlators in the free $O(N)$ theory and simply associate an extra factor of the $k_\perp$ momentum component flowing through the $\phi^2 \sigma$ vertex.  The reasoning is as follows.

Consider any correlator with some insertions of $\sigma$ in the interacting $O(N)$ theory, with no external $q_-$ flowing in through the $\sigma$ propagator.
We can write the $\sigma$ propagator in terms of its spectral function in mixed position/momentum space as
\be
\< \sigma(x^+, q_-, q_\perp) \sigma(0, -q_-, -q_\perp)\> &=& \int d\mu^2 dq_+ \frac{\mu e^{i q_+ x^+}}{2q_+ q_- - q_\perp^2 - \mu^2+i \epsilon} .
\ee
The relevant deformation $\sigma$ is integrated along all $x^-$ and therefore has $q_-=0$.  With $q_-=0$, it is clear that there are no poles as a function of $q_+$, so if $x^+ \ne 0$, then the propagator vanishes; the only possible contribution is proportional to $\delta(x^+)$.  We can compute the coefficient of the $\delta(x^+)$ function by integrating,
\be
\int dx^+ \int_0^\Lambda d\mu^2 dq_+ \frac{\mu e^{i q_+ x^+}}{ - q_\perp^2 - \mu^2+i \epsilon} = - \int_0^\Lambda \frac{d\mu^2 \mu}{q_\perp^2 + \mu^2} = \pi |q_\perp|,
\ee
where we have subtracted off a UV divergence $= -2 \Lambda$.  Therefore, 
\be
\< \sigma(x^+,0, q_\perp) \sigma(0, 0, -q_\perp)\> \cong \delta(x^+) \pi |q_\perp| .
\ee
In our prescription, we integrate over $x^+$, so we just pick up the contribution $\pi |q_\perp|$.  A 1PI diagram in the interacting theory, with the external $\sigma$ operators, is therefore reduced to diagram in the free theory, without them, times a factor of $\pi |q_\perp|$.


\section{Discussion and Future Directions}
\label{sec:discussion}

Our main result in this paper is the prescription (\ref{eq:Heff}) for an effective lightcone Hamiltonian $H_{\rm eff}$ that incorporates the effects of integrating out zero modes.  The proposed $H_\eff$ was defined directly in terms of CFT correlation functions, without reference to an underlying Lagrangian.  One can regard $H_\eff$ as the result of integrating out the zero modes before taking the LC limit.

We have shown how to apply our prescription in several examples, but there is more work to do to understand how $H_{\rm eff}$ behaves in various specific theories.  However, it seems very plausible that many theories will not have any contribution to $H_{\rm eff}$ beyond the vacuum energy and the renormalization of bare parameters.  The reason is that, as we saw in section \ref{sec:DefinitionZeroModes}, LC quantization only discards zero mode contributions between operators with definite relations in their dimensions.  Such relations arise in free theories and infinite $N$ theories, and often in  integrable theories, but rarely in generic non-integrable theories.

Of course, free theories  are a particularly important class, since their CFT data is readily available. It is likely that the types of arguments employed in sections \ref{sec:Plants}, \ref{sec:phi4}, and  \ref{sec:ON}  could be generalized to a large class of perturbative theories.  The goal would be to show that perturbative deformations  lead to an $H_{\rm eff}$ whose deviations from the naive LC Hamiltonian can be completely absorbed by a shift in the bare parameters of the theory, plus  terms like the non-local fermion bilinear term in (\ref{eq:FermEff}) that result from integrating out non-dynamical fields.  At a diagrammatic level, this seems plausible because diagrams contributing to $H_{\rm eff}$ must have very special configurations\footnote{The so-called plant diagrams, such as figure~\ref{fig:PlantDiagram}, are the obvious way to satisfy such constraints, and they only renormalize bare parameters.   At least in perturbation theory, a general rule for which non-plant diagrams (if any) contribute to $H_{\rm eff}$ seems attainable.  Moreover, beyond the free limit anomalous dimensions will remove integer relations between operator dimensions.  Aside from operators with protected dimensions, we therefore expect that deviations from the naive LC Hamiltonian will vanish.  Perturbative CFTs may be a useful concrete arena in which to try to study this expectation.}
 in order to produce $\delta(x^+)$ functions in LC time.

In general, it may be more natural to study $H_{\rm eff}$ using CFT data.  Unfortunately, the OPE is not so well-behaved in momentum space.  Most of our analyses have been either fully in momentum space or else in mixed position/momentum space.  It would be useful to understand if they can nevertheless be reformulated in terms of a convergent OPE.

Our prescription for constructing $H_\eff$ allows one to match LC and ET calculations to all orders in the deformation parameters. However, it is still possible for there to be additional non-perturbative effects  which are missed by this construction, as we demonstrated in the case for 2d $\phi^4$ theory in section~\ref{sec:phi4}. While one might hope that the perturbative data is sufficient to determine these non-perturbative effects either quantitatively or at least qualitatively (for instance, whether these effects can simply be absorbed into a shift in bare parameters), it remains unclear if this is generally the case. We plan to consider these effects in more detail in future work, which must be understood in order to develop a fully non-perturbative prescription for integrating out zero modes.

Looking ahead,  perhaps our most important conclusion is that RG flows originating from $\Ncal=4$ SYM at infinite $N$ can be investigated using lightcone Hamiltonian truncation methods.  If our expectations about $H_{\rm eff}$ bear out, then at least at finite 't Hooft coupling $\lambda$ the only contribution to $H_{\rm eff}$ arises from integrating out the non-dynamical component of fermion fields on the LC.  The only other data that is needed for LC Hamiltonian truncation is the spectrum of operators, which are known, and three-point functions of single-trace operators, which have been obtained up to three loops \cite{Basso:2015zoa,Basso:2015eqa}.\footnote{Dealing with terms generated by integrating out the fermion fields remains a non-trivial complication.  One strategy for avoiding this problem is to compute matrix elements of the supercharge, and use the superconformal algebra to compute matrix elements of the Hamiltonian.}   So it may be possible to obtain the spectra of large $N$ confining gauge theories in practice by perturbing $\Ncal=4$ SYM theory.

\section*{Acknowledgments}    

We would like to thank David Andriot, Marc Gillioz, Zuhair Khandker, Markus Luty, Slava Rychkov, and Andreas Stergiou for valuable discussions, as well as Marco Serone and Gabriele Spada for sharing with us perturbative coefficients of the $\phi^4$ theory. ALF,  EK, LGV, and MW were supported in part by the US Department of Energy Office
of Science under Award Number DE-SC0015845.  JK was supported in
part by NSF grant PHY-1454083.  ALF, JK, EK, LGV, and MW were also supported in part by the
Simons Collaboration Grant on the Non-Perturbative Bootstrap. We would also like to thank the ICTP South American Institute for Fundamental Research and the Institut des Hautes \'{E}tudes Scientifiques and the workshop on ``Hamiltonian methods in strongly coupled Quantum Field Theory''  for hospitality while this work was completed.

\begin{appendices}

\section{Hamiltonian Matrix Elements from Correlation Functions}
\label{app:MEfromC}
 
In this appendix, we consider the standard method for computing Hamiltonian matrix elements from CFT three-point functions. First, we briefly review the use of $i\ep$ prescriptions in Lorentzian correlation functions. Using this prescription, we then explicitly compute Hamiltonian matrix elements for the example of $d=2$, demonstrating that a large set of matrix elements vanish for free or large $N$ theories.


\subsection{Lorentzian Correlators and the $i\ep$ Prescription}

In standard conformal truncation, Hamiltonian matrix elements are given by Fourier transforms of Lorentzian correlation functions. These correlators can in turn be defined as analytic continuations of Euclidean correlation functions. However, there are ambiguities in this analytic continuation, depending on our choice of contour. These different contours correspond to different orderings of the operators in the resulting Lorentzian correlator.

Fortunately, we can use a simple $i\epsilon$ prescription to fix a particular choice of contour, and thus a particular operator ordering (for a nice review, see \cite{Hartman:2015lfa}). In general, we can obtain a specific ordering for Lorentzian correlators with the following prescription
\be
\<\Ocal_1(t_1,\vec{x}_1) \cdots \Ocal_n(t_n,\vec{x}_n)\> = \lim_{\ep_i\ra0} \<\Ocal_1(t_1-i\ep_1,\vec{x}_1) \cdots \Ocal_n(t_n-i\ep_n,\vec{x}_n)\>,
\ee
where the limit is taken with $\ep_1 > \cdots > \ep_n$ (such that $\Ocal_1$ is the leftmost operator and $\Ocal_n$ the rightmost). In terms of the lightcone coordinates $x^\pm \equiv \fr{1}{\sqrt{2}}(t\pm x)$, this prescription becomes
\be
x^\pm_i \ra x^\pm_i - i \ep_i.
\ee

As a simple example, let's consider a two-point function in $d=2$. Using this $i\ep$ prescription, we find the two orderings
\be
\begin{split}
\<\Ocal(x_1) \Ocal(x_2)\> &= \fr{1}{\left((x_{12}^+ - i\ep)(x_{12}^- - i\ep)\right)^\De}, \\
\<\Ocal(x_2)\Ocal(x_1)\> &= \fr{1}{\left((x_{12}^+ + i\ep)(x_{12}^- + i\ep)\right)^\De}.
\end{split}
\ee
We can also use these two expressions to obtain the time-ordered correlation function
\be
\begin{split}
\<\Tcal\{\Ocal(x_1) \Ocal(x_2)\}\> = \fr{1}{\left((x_{12}^+ - i\ep \, \sgn(t_{12}))(x_{12}^- - i\ep \, \sgn(t_{12}))\right)^\De} = \fr{1}{(x_{12}^+ x_{12}^- - i\epsilon)^\De}.
\end{split}
\ee

We could also consider ordering operators with respect to the lightcone time $x^+$, rather than the standard time $t$. However, this $i\epsilon$ prescription is only necessary if the two operators are timelike separated. In other words, both orderings are equivalent at spacelike separation, since the operators commute due to causality,
\be
\comm{\Ocal(x_1)}{\Ocal(x_2)} = 0 \qquad (x_{12}^2 < 0).
\ee
At timelike separation $\sgn(x^+) = \sgn(t)$, which means that lightcone time-ordering is actually the \emph{same} as standard time-ordering in causal theories.


\subsection{Computing Matrix Elements}

For a general CFT deformed by a relevant operator $\Ocal_R$, the resulting Hamiltonian matrix elements are given by
\be
\<\Ocal,\vec{P},\mu|V|\Ocal',\vec{P}',\mu'\> \equiv \lambda \int d^dx_1 \, d^{d-1}x_2 \, d^dx_3 \, e^{i(P\cdot x_1 - P'\cdot x_3)} \<\Ocal(x_1)\Ocal_R(x_2)\Ocal'(x_3)\>.
\ee
The three-point function in the integrand is a Wightman function with a specific ordering. We therefore need to use the appropriate $i\ep$ prescription to enforce this ordering. For example, in $d=2$ this correlator takes the form
\be
\begin{split}
\<\Ocal(x_1) \Ocal_R(x_2) \Ocal'(x_3)\> &= \fr{C_{\Ocal\Ocal'\Ocal_R}}{(x_{12}^- - i\ep)^{h+h_R-h'} (x_{23}^- - i\ep)^{h'+h_R-h} (x_{13}^- - i\ep)^{h+h'-h_R}} \\
&\qquad \times \, \fr{1}{(x_{12}^+ - i\ep)^{\bar{h}+\bar{h}_R - \bar{h}'} (x_{23}^+ - i\ep)^{\bar{h}'+\bar{h}_R-\bar{h}} (x_{13}^+ - i\ep)^{\bar{h}+\bar{h}'-\bar{h}_R}},
\end{split}
\ee
where $h$ and $\bar{h}$ are the (anti)holomorphic dimensions
\be
h \equiv \De + \ell, \quad \bar{h} \equiv \De - \ell.
\ee

\begin{figure}[t!]
\centering
\includegraphics[width=.4\linewidth]{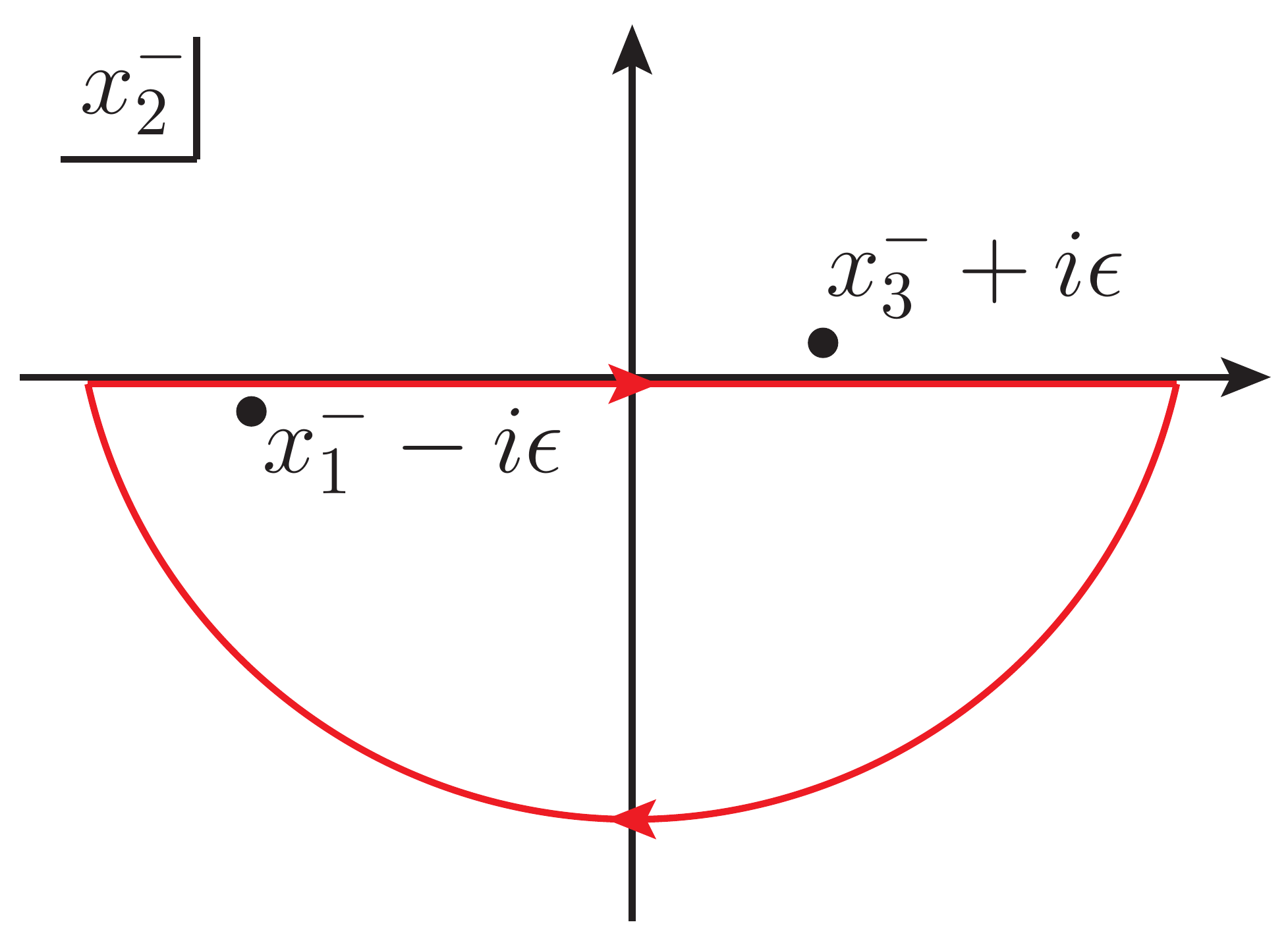}
\caption{Integration contour for evaluating Hamiltonian matrix elements. The $i\ep$ prescription forces the two branch points to lie on opposite sides of the real axis. If the scaling dimensions are related by an integer, such that $h' = h + h_R + n$, one of the branch points becomes a zero, and the integral vanishes.}
\label{fig:PrescriptionLC}
\end{figure}

For this 2d example, let's specifically focus on the integral over the insertion of the relevant deformation, $x_2^-$. This integral takes the form
\be
I(x_1^-,x_3^-) \equiv \int dx_2^- \, \fr{1}{(x_{12}^- - i\ep)^{h+h_R-h'} (x_{23}^- - i\ep)^{h'+h_R-h}}.
\ee
Looking at the integrand, we see that there are two branch points, where $\Ocal_R$ collides with one of the other operators. Due to our $i\epsilon$ prescription, one of these branch points is located in the upper half plane and the other is in the lower half plane, as shown in figure~\ref{fig:PrescriptionLC}. So long as the relevant deformation has $\De_R > \fr{d}{2}$, we can then evaluate this integral by closing the contour on either side, leading to the result
\be
I(x_1^-,x_3^-) = \fr{2\pi i \G(2h_R-1)}{\G(h+h_R-h')\G(h'+h_R-h) (x_{13} - i\ep)^{2h_R-1}}.
\ee
However, if the holomorphic dimensions are related by an integer, such that
\be
h' = h + h_R + n,
\ee
then one of the gamma functions in the denominator is singular, such that this expression vanishes.

This behavior is easy to understand from the associated integration contour. When the dimensions are related by an integer, the integrand takes the form
\be
\fr{1}{(x_{12}^- - i\ep)^{h+h_R-h'} (x_{23}^- - i\ep)^{h'+h_R-h}} = \fr{(x_{12}^- - i\ep)^n}{(x_{23}^- - i\ep)^{2h_R+n}}.
\ee
In this case, there is no longer a branch point at $x_1^-$, such that if we close the contour in the lower half plane the integral (and the resulting Hamiltonian matrix element) vanishes,
\be
\boxed{\<\Ocal,P_-,\mu|V_\LC|\Ocal',P'_-,\mu'\> = 0 \qquad (h' = h + h_R + n).}
\ee

For theories at large $N$, this relation between dimensions precisely corresponds to the case where one of the operators is a double-trace operator built from the other two,
\be
\Ocal' = [\Ocal\Ocal_R]_{n,\bar{n}} \equiv \Ocal \lrpar_-^n \lrpar_+^{\bar{n}} \Ocal_R.
\ee
We thus see that \emph{all} matrix elements mixing single-trace operators with double-trace operators vanish at infinite $N$,
\be
\<\Ocal,P|V_\LC|[\Ocal\Ocal_R]_{n,\bar{n}},P'\> = 0 \qquad (N\ra\infty).
\ee
While we've focused on the case of $d=2$ for the sake of simplicity, this result can be generalized to arbitrary $d$.


\section{Details of the Bulk Model Old-Fashioned Perturbation Theory}
\label{app:OFPTdetails}

Here we will go through the details of the Old-Fashioned Perturbation Theory (OFPT) computation, referenced in section \ref{sec:MissingZero}, of the energy eigenvalues in the bulk toy model at second order.  We want to evaluate the contributions at  second order, (\ref{eq:OFPT2nd}), from double-trace operators.  

The relevant operator is a deformation to the CFT described by the interacting theory in the bulk with a quartic coupling $\sim g_4 \phi^4$.  To take the effect of the bulk interaction into account, we will also work to first order in an expansion at small $g_4$.  The computation will be easiest to do if we first sum over double-trace operators in a bulk Fock space basis; once we have the answer, it will be straightforward to interpret the result as a sum over primary operators.  

 To begin, we write the double-trace states corrected by the $g_4 \phi^4$ coupling in the bulk at leading order:
\be
| p_1, p_2 \> ^{(1)} &=& |p_1, p_2 \> + \int d^2 \tilde{p}_1 d^2 \tilde{p}_2 \frac{\< \tilde{p}_1 \tilde{p}_2 | \int dz dx_- \sqrt{g} g_4 \phi^4 | p_1 , p_2 \>}{\frac{\tilde{\mu}_1^2}{p_{1-}} + \frac{\tilde{\mu}_2^2}{p_{2-}} - \frac{\mu_1^2}{p_{1-}} - \frac{\mu_2^2}{p_{2-}} } | \tilde{p_1} \tilde{p}_2\> .
\label{eq:FirstOrderBulkOFPT}
\ee
States without an $^{(n)}$ superscript are eigenstates in the absence of both the bulk interaction $g_4 \phi^4$ and the boundary deformation $V$.  The above expression is the expansion in $g_4 \phi^4$ and should not be confused with the expansion in the relevant deformation $V \sim \lambda \Ocal(q_-)$; we will perform two separate OFPT expansions, one in $g_4$ (to first order), and one in $V$ (to second order).  To avoid clutter, from now on we will use the abbreviation
\be
\int g_4 \phi^4 \cong \int dz dx_- \sqrt{g} g_4 \phi^4.
\ee
The denominator in (\ref{eq:FirstOrderBulkOFPT})  is the energy denominator, and $\mu^2$ is the mass-squared of each state. The integral $d^2 \tilde{p}_1 d^2 \tilde{p}_2$ can equally well be thought of as an integral over $\tilde{\mu}_1, \tilde{\mu}_2, \tilde{p}_{1-}, \tilde{p}_{2-}$.  

Next, we return to the second order OFPT term in the expansion in the relevant deformation $V$:
\be
\delta P^+ \equiv \Big[ \< \Ocal, p, \mu | H | \Ocal, p', \mu'\> \Big]^{(2)} \supset \int d^2 p_1 d^2 p_2 \frac{\< p | V | p_1 p_2 \>^{(1)} {}^{(1)}\< p_1, p_2 | V | p' \> }{\frac{\mu^2}{p_-}- \frac{\mu^2_1}{p_{1-}} - \frac{\mu_2^2}{p_{2-}}} + (p \leftrightarrow p') .
\label{eq:Pert11}
\ee
We have written $\supset$ instead of $=$ above because we are just considering the contribution from the double-trace states at this order.  
Substituting the expression for $|p_1 p_2\>^{(1)}$ into $\delta P^+$ above, we need the following overlaps:
\be
\< p | \int \lambda \CO | p_1, p_2\>^{(0)} &=& \lambda  \mu_2^\nu \delta^2(p-p_1)\delta(p_{2-}-q_-) + {\rm sym} , \quad \nu \equiv \Delta-1, 
\ee
and
\begin{equation}
\Ical \equiv \< p | V  \int d^2 \tilde{p}_1 d^2 \tilde{p}_2 \frac{\< \tilde{p}_1 \tilde{p}_2 | \int g_4 \phi^4 | p_1 , p_2 \>}{\frac{\tilde{\mu}_1^2}{p_{1-}} + \frac{\tilde{\mu}_2^2}{p_{2-}} - \frac{\mu_1^2}{p_{1-}} - \frac{\mu_2^2}{p_{2-}} } | \tilde{p_1} \tilde{p}_2\> 
 = \lambda \int \frac{d\tilde{\mu}_2^2 \tilde{\mu}_2^\nu}{q_-} \frac{\< p, q_-, \tilde{\mu}_2 | \int g_4 \phi^4 | p_1 , p_2 \>}{\frac{\mu^2}{p_-} + \frac{\tilde{\mu}_2^2}{q_-} - \frac{\mu_1^2}{p_{1-}} - \frac{\mu_2^2}{p_{2-}} } . 
\end{equation}
The matrix element of $g_4 \phi^4$  on the RHS above can be evaluated using standard methods for Witten diagrams:
\be
\Ical &=& \lambda \int \frac{d \tilde{\mu}_2^2 \tilde{\mu}_2^\nu}{q_-} \frac{\Acal(\mu, \mu_1, \mu_2, \tilde{\mu}_2)}{\frac{\mu^2}{p_-} + \frac{\tilde{\mu}_2^2}{q_-} - \frac{\mu_1^2}{p_{1-}} - \frac{\mu_2^2}{p_{2-}} } , \\
\Acal(\mu, \mu_1, \mu_2, \tilde{\mu}_2)&=& \< p, q_-, \tilde{\mu}_2 | \int g_4 \phi^4 | p_1 , p_2 \> = g_4 \int_0^\infty z dz J_\nu(\mu_1 z) J_\nu(\mu_2 z) J_\nu(\tilde{\mu}_2 z) J_\nu(\mu z) . \nn
\ee
 Putting everything together, one finds
 \be
 \delta P^+ \supset \int \frac{d \mu_2^2 d\tilde{\mu}_2^2  \mu_2^\nu \tilde{\mu}_2^\nu}{(\mu^2 (1- \frac{q_-}{p_-}) - \mu_2^2)( \tilde{\mu}_2^2 - \mu_2^2)} \Acal(\mu,\mu, \mu_2, \tilde{\mu}_2) .
 \ee
The integral is over the mass-squareds of the Fock space modes, but we are interested in the contribution from the mass-squareds of the full double-trace states.  To change variables, we start with the mass-squared $M^2$ of the double-trace state:
\be
M^2 = (p_1+ p_2)^2 = \mu_1^2 + \mu_2^2  +  \mu_2^2 \frac{p_{1-}}{p_{2-}} + \mu_2^2 \frac{p_{2-}}{p_{1-}}.
\ee
In the evaluation of $\delta P^+$, $\delta$ functions set $p^\mu = p_1^\mu$ and $p_{2-} = q_-$, so
\be
M^2 \cong \mu^2 \left( 1 + \frac{q_-}{p_-} \right) + \mu_2^2 \left(1+ \frac{p_-}{q_-} \right). 
\ee
In the limit of small $q_- \ll p_- $ and large $M^2 \gg \mu^2$ (with fixed external momentum $p$ and $\mu^2$), we therefore have the relation 
\be
\mu_2^2 \approx M^2 q_-/p_-.
\ee
Finally, we obtain the relation described in the text:
\begin{equation}
\delta P^+ \sim \int d\tilde{\mu}_2^2 d M^2 \frac{q_-}{p_-}  \tilde{\mu}_2^\nu \left(M^2 \frac{q_-}{p_-} \right)^{\frac{\nu}{2}-2}\Acal(\mu, \mu, M^2 \frac{q_-}{p_-} , \tilde{\mu}_2), \qquad ( M^2 \gg \mu^2, \tilde{\mu}_2^2 \textrm{ and }  q_- \ll p_- ).
\end{equation}

We would like to understand how this result scales at large $M^2$ and small $q_-$ for general values of $\nu$.  This will be determined by the integral
\be
\delta P^+ \sim \int d \tilde{\mu}_2^2 \, \tilde{\mu}_2^\nu \left(M \sqrt{ \frac{q_-}{p_-}}\right)^{\nu-2} \Acal (\mu, \mu, M (q_-/p_-) ^{1/2} , \tilde{\mu}_2)
\ee
which in turn depends on the amplitude
\be
\Acal(\mu, \mu, \mu_2, \tilde{\mu}_2) = g_4 \int_0^\infty z dz \left[ J_\nu(\mu z) \right]^2 J_\nu(\mu_2 z) J_\nu(\tilde{\mu}_2 z) 
\ee
evaluated at large $\mu_2$.  We can immediately evaluate the $\tilde \mu_2$ integral since only a single Bessel function in the amplitude depends on this variable, yielding
\be
\label{eq:simplifiedlargeMdPplus}
\delta P^+ \sim g_4 \left(M \sqrt{ \frac{q_-}{p_-}}\right)^{\nu-2} \frac{2^\nu \Gamma(\nu + \frac{1}{2})}{\sqrt{\pi}} \int \frac{dz}{z^\nu} \left[ J_\nu(\mu z) \right]^2 J_\nu \left( z M \sqrt{\frac{q_-}{p_-}}  \right) 
\ee
Note that for $\nu > -\frac{1}{2}$ the integrand is convergent near $z = 0$.  If we expand at large $M$, we can simplify the Bessel function 
\be
J_\nu \left( z M \sqrt{\frac{q_-}{p_-}}  \right)  \approx \frac{\sqrt{\frac{2}{\pi }} \sin \left(\frac{1}{4} (-2 \pi  \nu +4 z M \sqrt{\frac{q_-}{p_-}} 
   +\pi )\right)}{\sqrt{z M \sqrt{\frac{q_-}{p_-} } } }
\ee
In this approximation, the remaining $z$ integral essentially becomes the large `energy' Fourier transform of 
\be
f(z) = z^{-\nu} \left[ J_\nu(\mu z) \right]^2
\ee
with respect to $z$.   It appears that its possible to directly evaluate this integral in terms of ${}_3 F_2$ hypergeometric functions (producing a final result that scales as $1/M^3$ for all $\nu$), but there is a better way to understand the large $M$ behavior.

At very large values of $M$, the Fourier transform will be dominated by the least analytic parts of $f(z)$.  Since $f$ is smooth at general values of $z$, this means that the transform will be dominated by the boundary of the region of integration, namely small $z$.  In this limit $f(z) \sim z^\nu$, so we can evaluate equation (\ref{eq:simplifiedlargeMdPplus}) simply by rescaling the integration variable, giving
\be
\delta P^+ \sim g_4 \frac{1}{\left(M \sqrt{ \frac{q_-}{p_-}} \right)^{3} } 
\ee
for all values of $\nu$ at large $M$, where we have neglected many numerical factors.

\section{SUSY Bulk Model}
\label{app:SUSY}
In this Section we analyze a large N model with local bulk Lagrangian in which the bulk profile can be solved exactly.
As discussed in Section \ref{sec:Problem}, finding the correct AdS vacuum profile in necessary to describe the correct IR QFT.

In particular, consider a ``supersymmetric'' bulk Lagrangian for a single scalar field, in terms of a superpotential 
$W(\phi)$,
\begin{align}
S &= N^2 \int d^{d+1} x \sqrt{-g} \frac{1}{2} \left[ z^2 \left(\partial_z \phi - \frac{1}{z} 
\frac{\partial W}{\partial \phi} \right)^2 + \frac{1}{2}z^2 (\partial_\mu \phi)^2 \right] 
\\ & \sim N^2 \int d^{d+1} x \sqrt{-g} \left[ \frac{1}{2} z^2 (\partial_z \phi)^2 - \,d \,W + \frac{1}{2} 
\left(\frac{\partial W}{\partial \phi}\right)^2 + \frac{1}{2} z^2 (\partial_\mu \phi)^2 \right] \,.
\end{align}
For definiteness, let us fix the form of the superpotential,
\begin{equation}
W = \frac{m}{2} \phi^2 - \frac{g}{3!} \phi^3 \,,
\label{eq:superpot}
\end{equation}
corresponding to a bulk mass $M^2 = m (d-m)$ and UV conformal dimension $\Delta = m$ or 
$\Delta = d-m$ for the boundary operator. In this paper we consider the second case, since deforming the CFT
by an operator with dimension $\frac{d}{2}<\Delta<d$ can be more directly treated with Conformal Truncation.
 
The exact background solution is found by solving $\partial_z \phi_{\rm cl} = \frac{1}{z} 
\left. \frac{\partial W}{\partial \phi} \right|_{\phi = \phi_{\rm cl}}$, which in case of the explicit superpotential 
\eqref{eq:superpot} gives
\begin{equation}
\phi_{\rm cl}(z) = \lambda \frac{z^m}{1+\frac{g}{2 m} \lambda z^m} \,
\label{eq:vev}
\end{equation}
which in the UV approaches
\begin{equation}
\label{eq:bdy}
\phi_{\rm cl} \stackrel{z\to 0}{\sim} \lambda z^m + o(z^{2 m}) \,.
\end{equation}
According to the standard AdS/CFT dictionary, the boundary condition \eqref{eq:bdy} is equivalent to deforming the boundary 
action by $\delta S = N \lambda \int \Ocal_\Delta$. Note that the power $z^{d-m}$ is absent in \eqref{eq:bdy}, so that 
$\langle \Ocal_\Delta \rangle = 0$. This is property holds for any form of the local superpotential $W$. 

In the IR, the bulk profile \eqref{eq:vev} flows to a constant, $\phi_{\rm cl} \stackrel{z\to \infty}{\sim} \frac{2 m}{g}$.
Expanding the potential around the asymptotic profile, the mass squared for the linearized spectrum of perturbations is 
$V''(\phi_{\rm cl}) = m(d +m)$. That corresponds to an IR CFT with spectral dimension $\Delta_{\rm IR} = d+m = 
2 d - \Delta$, which is irrelevant.

Using Conformal Truncation, it is possible to plugin \eqref{eq:bdy} into \eqref{eq:deltah} to compute the effective Hamiltonian 
for the spectrum of perturbations around $\phi_{\rm cl}$. That would give access to information on the full RG flow, 
for example via the spectral density of $\Ocal_\Delta$. That, however, lies outside the scope of the present work.


\section{Lightcone Truncation and the Infinite Momentum Limit}
\label{app:LCvsET}

In this appendix, we demonstrate that the matrix elements of the ``naive'' lightcone Hamiltonian, which are computed from CFT three-point functions, correspond to the infinite momentum limit of matrix elements of the more familiar equal-time Hamiltonian. While this result is perhaps not surprising, establishing this relation is an important step in justifying our prescription for the effective LC Hamiltonian.

To start, let's briefly review the structure of conformal truncation in ET and LC quantization. In both cases, the correction to the Hamiltonian density simply corresponds to a relevant local operator $\Ocal_R(x)$. The resulting ET Hamiltonian is given by integrating this relevant operator over a slice of fixed time $t$,
\be
H = H_0 + V_\ET , \qquad V_\ET \equiv \lambda \int d^{d-1} x \, \Ocal_R(t,\vec{x}),
\ee
while the LC Hamiltonian is obtained by integrating over a slice of fixed lightcone time $x^+$,
\be
P_+ = P_{+0} + V_\LC , \qquad V_\LC \equiv \lambda \int dx^- d^{d-2} x^\perp \, \Ocal_R(x^+,x^-,\vec{x}^\perp).
\ee
Here, $\vec{x}^\perp$ is the set of directions perpendicular to the lightcone directions $x^\pm \equiv \fr{1}{\sqrt{2}}(t \pm x)$.

We are specifically interested in computing the matrix elements of the invariant mass operator $M^2\equiv P^\mu P_\mu$. We can write this in terms of the ET Hamiltonian,
\be
M^2 = (H_0+V_\ET)^2 - \vec{P}^2 = M_0^2 + \acomm{H_0}{V_\ET} + V_\ET^2,
\label{eq:MsqET}
\ee
or in terms of the LC Hamiltonian,
\be
M^2 = \acomm{P_{+0} + V_\LC}{P_-} - \vec{P}_\perp^2 = M_0^2 + \acomm{P_-}{V_\LC}.
\label{eq:MsqLC}
\ee
For conformal truncation, we consider the matrix elements of $M^2$ in a basis of states with definite conformal Casimir $\Ccal$, spatial momentum $\vec{P}$, and invariant mass $\mu$,
\be
|\Ocal,\vec{P},\mu\> \equiv \int d^dx \, e^{-iP\cdot x} \Ocal(x)|0\>,
\label{eq:BasisDef}
\ee
where $\mu^2 \equiv P^2$ is the associated eigenvalue of the unperturbed mass operator $M_0^2$. The resulting ET basis states are labeled by the spatial momentum $\vec{P} = (P_x, \vec{P}_\perp)$, while the LC states are labeled by $\vec{P} = (P_-, \vec{P}_\perp)$. Note that these basis states are defined by Fourier transforming with respect to \emph{all} spacetime directions, regardless of quantization scheme.\footnote{As shown in~\cite{Gillioz:2016jnn}, this complete set of states in Minkowski space can be mapped to the more familiar radial quantization states via a combination of conformal transformations and Wick rotation.}

Because the physical mass-squared is Lorentz invariant, the eigenvalues of the full infinite-dimensional matrix constructed from $M^2$ are the same in both ET and LC quantization. However, the individual matrix elements in the two quantization schemes are generically different, due to the fact that the operator is acting on two different Hilbert spaces. If we truncate the two matrices by setting some $\Cmax$ and then diagonalize, we will therefore obtain two \emph{different} sets of eigenvalues. While these eigenvalues must converge to the same result as $\Cmax \ra \infty$, at any finite truncation level there will generically be some difference.

However, we now want to show that in the infinite momentum limit the individual matrix elements in ET quantization exactly match those of LC quantization. In other words, we can \emph{define} the LC Hamiltonian as the infinite momentum limit of the ET Hamiltonian. To do so, we first show that in the limit of infinite spatial momentum $(|P_x| \ra \infty)$, the matrix elements that are linear in $V$ match in the two quantization schemes,
\be
\lim_{|P_x| \rightarrow \infty} \< \Ocal,P_x,\mu|\acomm{H_0}{V_\ET}|\Ocal',P_x',\mu'\> = \<\Ocal,P_-,\mu|\acomm{P_-}{V_\LC}|\Ocal',P_-',\mu'\>.
\ee 
We then show that the matrix elements of $V_\ET^2$ vanish at infinite momentum,
\be
\lim_{P_x \rightarrow \infty} \<\Ocal,P_x,\mu|V_\ET^2|\Ocal',P_x',\mu'\>  = 0.
\ee

As a first step, consider the normalization of our basis states. Given the definition in eq.~\eqref{eq:BasisDef}, we find that the inner product is simply the Fourier transform of a CFT two-point function,
\be
\<\Ocal,\vec{P},\mu|\Ocal,\vec{P}',\mu'\> = \int d^dx_1 \, d^dx_2 \, e^{i(P\cdot x_1 - P'\cdot x_2)} \<\Ocal(x_1) \Ocal(x_2)\>.
\ee
Since the transverse momenta are conserved in both quantization schemes, we can specifically consider the reference frame with $\vec{P}_\perp = 0$, without loss of generality.

Depending on our choice of quantization scheme, we can then rewrite this inner product in the form
\be
\<\Ocal,\vec{P},\mu|\Ocal,\vec{P}',\mu'\> = 
\begin{cases}
2E (2\pi) \de(P_x - P_x') \, \de(\mu^2 - \mu^{\prime2}) \, \Ncal_\Ocal(P) & (\ET) \\
2P_- (2\pi) \de(P_- - P_-') \, \de(\mu^2 - \mu^{\prime2}) \, \Ncal_\Ocal(P) & (\LC)
\end{cases}
\ee
where, for simplicity, we've suppressed the overall delta functions for the transverse momenta. Note that both quantization schemes have the \emph{same} overall normalization factor
\be
\Ncal_\Ocal(P) \equiv \int d^dx \, e^{iP\cdot x} \<\Ocal(x) \Ocal(0)\>.
\ee
Let's look at this normalization factor more carefully. Because we've set $\vec{P}_\perp = 0$, this function can only depend on $\mu$ and $P_x$ (or equivalently $\mu$ and $P_-$). However, if we organize our basis into eigenstates of the operator $J_{+-}$, which generates boosts in the $t$-$x$ plane, then we can complete fix the $P_x$-dependence, obtaining the general expression
\be
\Ncal_\Ocal(P) = P_-^{2m} f_\Ocal(\mu),
\ee
where $m$ is the boost eigenvalue of $\Ocal$.

Turning to the mass-squared operator, we can write the matrix elements in a somewhat similar form,
\be
\<\Ocal,\vec{P},\mu|M^2|\Ocal',\vec{P}',\mu'\> = 
\begin{cases}
\sqrt{4EE'} (2\pi) \de(P_x - P_x') \, \Mcal_{\Ocal\Ocal'}^{(\ET)}(P,P') \\
2P_- (2\pi) \de(P_- - P_-') \, \Mcal_{\Ocal\Ocal'}^{(\LC)}(P,P')
\end{cases}
\ee
where we've again suppressed any $\vec{P}_\perp$ delta functions. In ET quantization, there are three distinct contributions to these matrix elements: the original CFT term $M_0^2$, the linear correction $\acomm{H_0}{V_\ET}$, and the quadratic correction $V_\ET^2$. Focusing first on the linear term, we can write the properly normalized ET matrix element as the Fourier transform of a CFT three-point function,
\be
\de\Mcal_{\Ocal\Ocal'}^{(\ET)}(P,P') = \fr{\lambda(E+E')}{\sqrt{4EE'\Ncal_\Ocal(P)\Ncal_{\Ocal'}(P')}} \int d^dx \, d^dx' \, e^{i(P\cdot x - P'\cdot x')} \<\Ocal(x) \Ocal_R(0) \Ocal'(x')\>.
\ee

Similar to the inner product, we can fix the $P_x$-dependence of this matrix element by using the transformation of the operators under $J_{+-}$, obtaining
\be
\de\Mcal_{\Ocal\Ocal'}^{(\ET)}(P,P') = \fr{\lambda(E+E')}{\sqrt{4EE'\Ncal_\Ocal(P)\Ncal_{\Ocal'}(P')}} \, P_-^m P_-^{\prime m'} g\big(\mu,\mu',\tfr{P_-'}{P_-}\big).
\ee
However, the overall scaling with respect to boosts is precisely cancelled by the normalization factors, reducing this to the somewhat simpler expression
\be
\de\Mcal_{\Ocal\Ocal'}^{(\ET)}(P,P') = \fr{\lambda(E+E')}{\sqrt{4EE'f_\Ocal(\mu)f_{\Ocal'}(\mu')}} \, g\big(\mu,\mu',\tfr{P_-'}{P_-}\big).
\ee
If we now take the limit $P_x \ra -\infty$,\footnote{The direction of this limit does not change the final result, but taking $P_x \ra -\infty$ is the natural choice in order to obtain slices of fixed $x^+$ (rather than $x^-$).} we find that the matrix element reduces to the $P_x$-independent expression
\be
\lim_{P_x \ra -\infty} \de\Mcal_{\Ocal\Ocal'}^{(\ET)}(P,P') = \fr{\lambda \, g(\mu,\mu')}{\sqrt{f_\Ocal(\mu)f_{\Ocal'}(\mu')}}.
\ee
The resulting expression \emph{exactly} matches the linear correction to the LC matrix elements,
\be
\begin{split}
\de\Mcal_{\Ocal\Ocal'}^{(\LC)}(P,P') &= \fr{\lambda}{\sqrt{\Ncal_\Ocal(P)\Ncal_{\Ocal'}(P')}} \int d^dx \, d^dx' \, e^{i(P\cdot x - P'\cdot x')} \<\Ocal(x) \Ocal_R(0) \Ocal'(x')\> \\
&= \fr{\lambda \, g(\mu,\mu')}{\sqrt{f_\Ocal(\mu)f_{\Ocal'}(\mu')}}.
\end{split}
\label{eq:LCSF}
\ee

We therefore see that the LC matrix elements simply correspond to the infinite momentum limit of ET matrix elements,
\be
\boxed{\lim_{|P_x| \ra \infty} \Mcal_{\Ocal\Ocal'}^{(\ET)}(\mu,\mu',P_x) = \Mcal_{\Ocal\Ocal'}^{(\LC)}(\mu,\mu').}
\ee
One interesting consequence of this relation is the observation that LC matrix elements are actually independent of the choice of reference frame, as all $P_-$-dependence cancels out. Lightcone truncation is thus the natural framework for making \emph{both} unitarity and Lorentz invariance manifest.

To complete this argument, though, we also need to establish that the matrix element contributions from $V_{\rm ET}^2$ vanish at infinite momentum,
\be
\lim_{P_x \rightarrow \infty} \< \Ocal , P_x, \mu |  V_{\rm ET}^2  | \Ocal', P_x', \mu'\>  = 0.
\ee
To see this, consider inserting a complete set of states,
\benn
\<\Ocal, P_x, \mu | V_{\rm ET}^2 |\Ocal,' P_x', \mu'\> \sim \sum_\psi \int dE \< \Ocal, P_x, \mu | V_{\rm ET} |\psi, P_x , E\> \< \psi, P_x, E |  V_{\rm ET} |\Ocal', P_x', \mu'\>,
\eenn
where we have labeled the intermediate states by their energy $E$ and spatial momentum $P_x$.  
The range of integration of $E$ is restricted to $E \ge P_x$, since the invariant mass-squared must be positive, but is also restricted to $ E \le \sqrt{P_x^2 + \Lambda^2} \approx P_x + \frac{\Lambda^2}{2P_x}$ in the presence of a UV cutoff, which we generically must introduce to even define the matrix elements of $V_\ET$. The range of integration for $E$ therefore vanishes as $P_x \ra \infty$.  Equivalently, the integral could instead be written in terms of the invariant mass $\mu_\psi^2$ of the intermediate state, in which case we obtain an explicit suppression factor of $1/\sqrt{P_x^2+ \mu_\psi^2}$ in the integration measure. The integral therefore vanishes at infinite momentum, since both the range of integration and the matrix elements in the integrand remain finite in the $P_x \rightarrow \infty $ limit.\footnote{Note that we can take the $P_x \rightarrow \infty$ limit after performing the $d \mu_\psi^2$ integration, so the subtleties that we saw in evaluating the Dyson series at infinite momentum do not arise here.}

\end{appendices}

\newpage
 
\bibliography{ConfDomBib}

\end{document}